\documentclass{aa}
\pdfoutput=1 
\usepackage{graphicx}
\usepackage{txfonts}

\begin{document}

   \title{2D non-LTE modelling of a filament observed in the H$\alpha$ line with the DST/IBIS spectropolarimeter}
   \titlerunning{2D non-LTE modelling of a filament observed in the H$\alpha$}

   \author{P. Schwartz
          \inst{1}
          \and
          S. Gun\'ar\inst{2}
          \and
          J. M. Jenkins\inst{3}
          \and
          D. M. Long\inst{3}
          \and
          P. Heinzel\inst{2}
          \and
          D. P. Choudhary\inst{4}
          }

   \institute{Astronomical Institute of Slovak Academy of Sciences, 05960 Tatransk\'{a} Lomnica,
              Slovak Republic\\
              \email{pschwartz@astro.sk}
         \and
        Astronomical Institute, The Czech Academy of Sciences, 25165 Ond\v{r}ejov, Czech Republic
         \and
        UCL-Mullard Space Science Laboratory, Holmbury St. Mary, Dorking, Surrey, RH5 6NT, UK
         \and
        Department of Physics \& Astronomy, California State University, Northridge, CA 91330-8268, USA}

   \date{Received XXX XX, XXXX; accepted XXX XX, XXXX}


  \abstract
      {We study a fragment of a large quiescent filament observed on May 29, 2017 by the Interferometric
        BIdimensional Spectropolarimeter
     (IBIS) mounted at the Dunn Solar Telescope. We focus on its quiescent stage prior to its eruption.}
    {We analyse the spectral observations obtained in the H$\alpha$ line to derive the thermodynamic properties of the
     plasma of the observed fragment of the filament.}
    {We used a 2D filament model employing radiative transfer computations under conditions that depart from the local
      thermodynamic equilibrium. We employed a forward modelling technique in which we used the 2D model to produce
      synthetic H$\alpha$ line profiles that we compared with the observations. We then found the set of model input parameters, 
      which produces synthetic spectra with the best agreement with observations.}
    {Our analysis shows that one part of the observed fragment of the filament is cooler, denser, 
     and more dynamic than its other part 
     that is hotter, less dense, and more quiescent. The derived temperatures in the first part range from
     $6,000$\,K to $10,000$\,K and in the latter part from $11,000$\,K to $14,000$\,K. The gas pressure is 
     $0.2$\,--\,$0.4\,\mathrm{dyn/cm}^{2}$ in
     the first part and around $0.15\,\mathrm{dyn/cm}^{2}$ in the latter part. The more dynamic nature of the
     first part is characterised by the
     line-of-sight velocities with absolute values of $6$\,--\,$7$\,km/s and microturbulent velocities of $8$\,--\,$9$\,km/s. On the 
     other hand, the latter part exhibits line-of-sight velocities with absolute values $0$\,--\,$2.5$\,km/s and microturbulent
     velocities of $4$\,--\,$6$\,km/s.}
   {}


   \maketitle
%
%
\section{Introduction}\label{s:intro}
%
%
\begin{figure*}
\resizebox{\textwidth}{!}{
\includegraphics{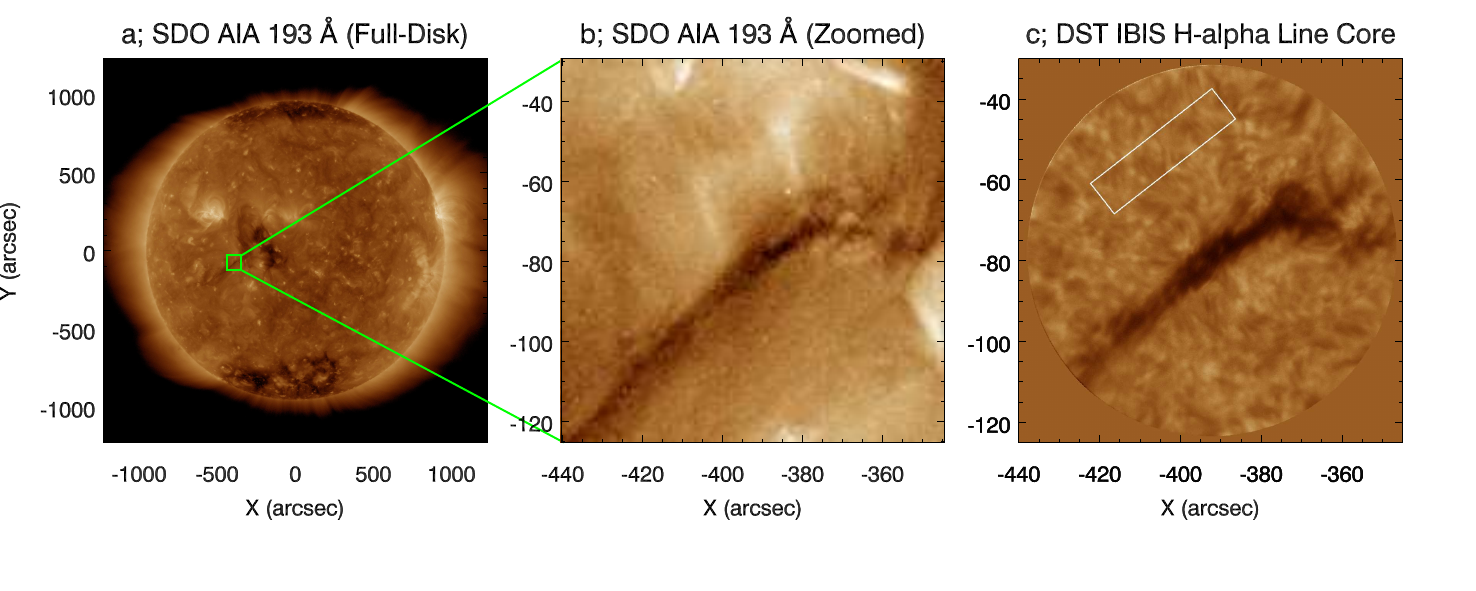}}\\ 
\parbox{0.33\textwidth}{
\resizebox{0.33\textwidth}{!}{
\includegraphics{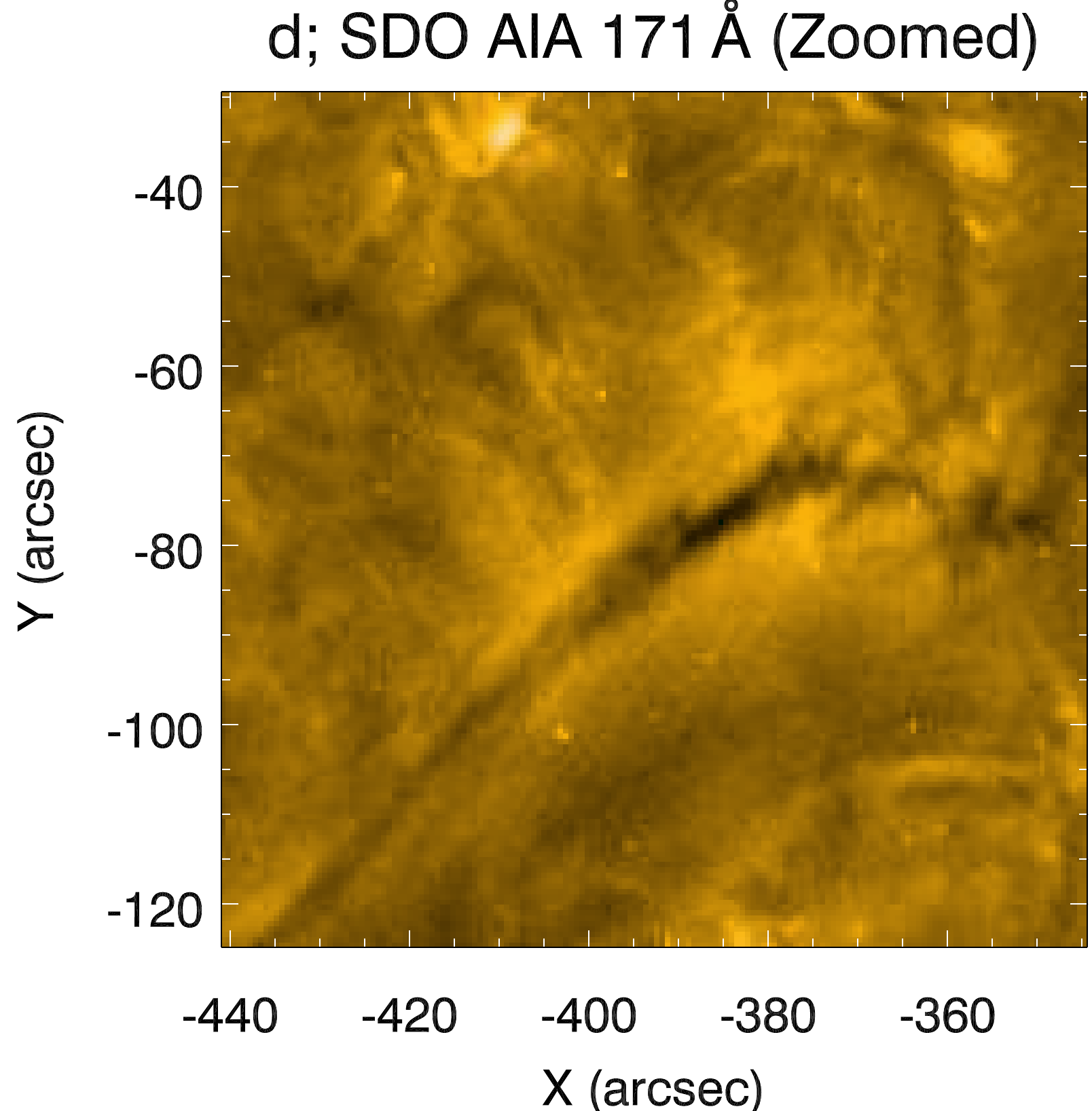}}}\ 
\parbox{0.33\textwidth}{
\resizebox{0.33\textwidth}{!}{
\includegraphics{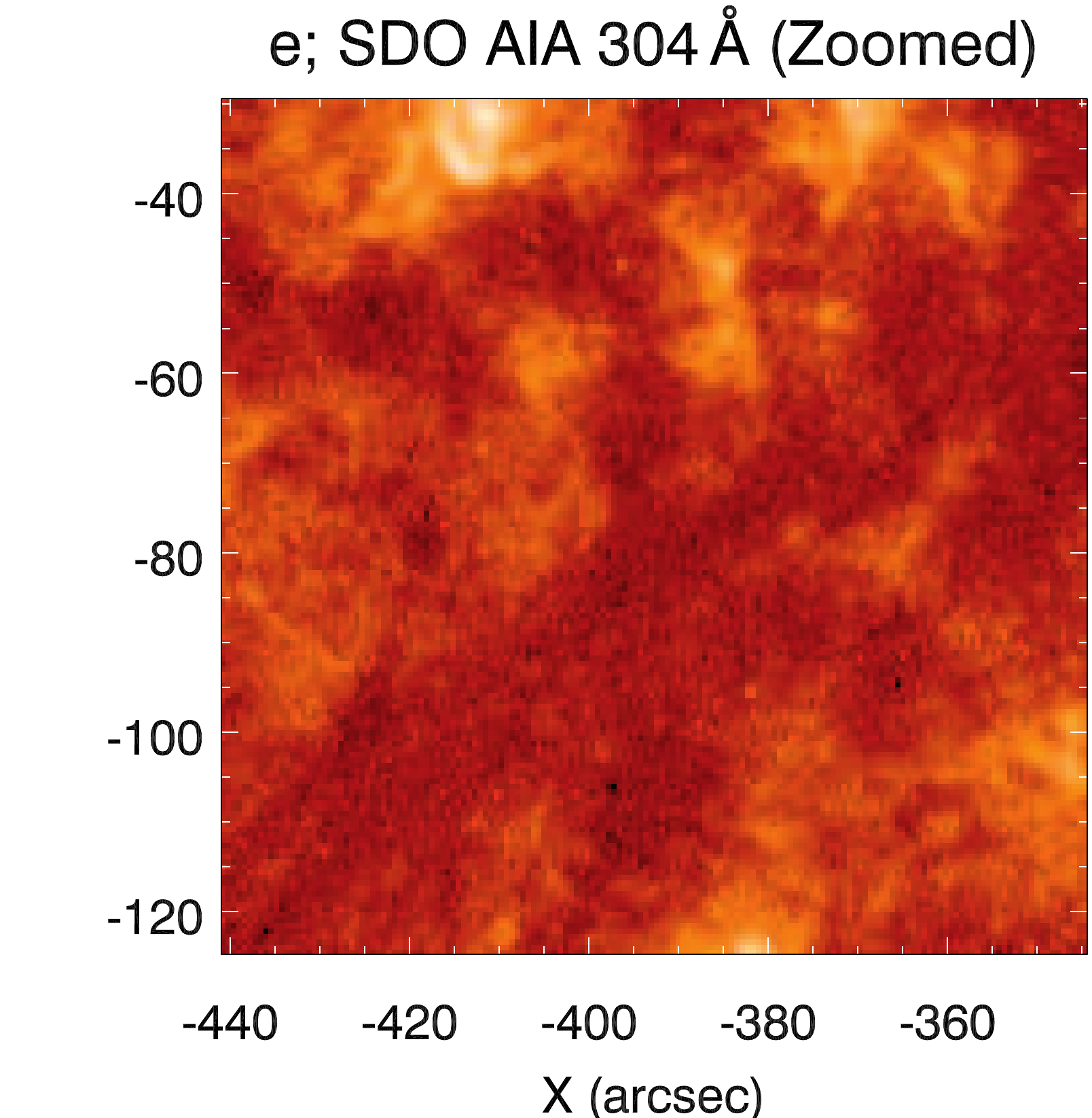}}}\ 
\parbox{0.33\textwidth}{
\resizebox{0.33\textwidth}{!}{
\includegraphics{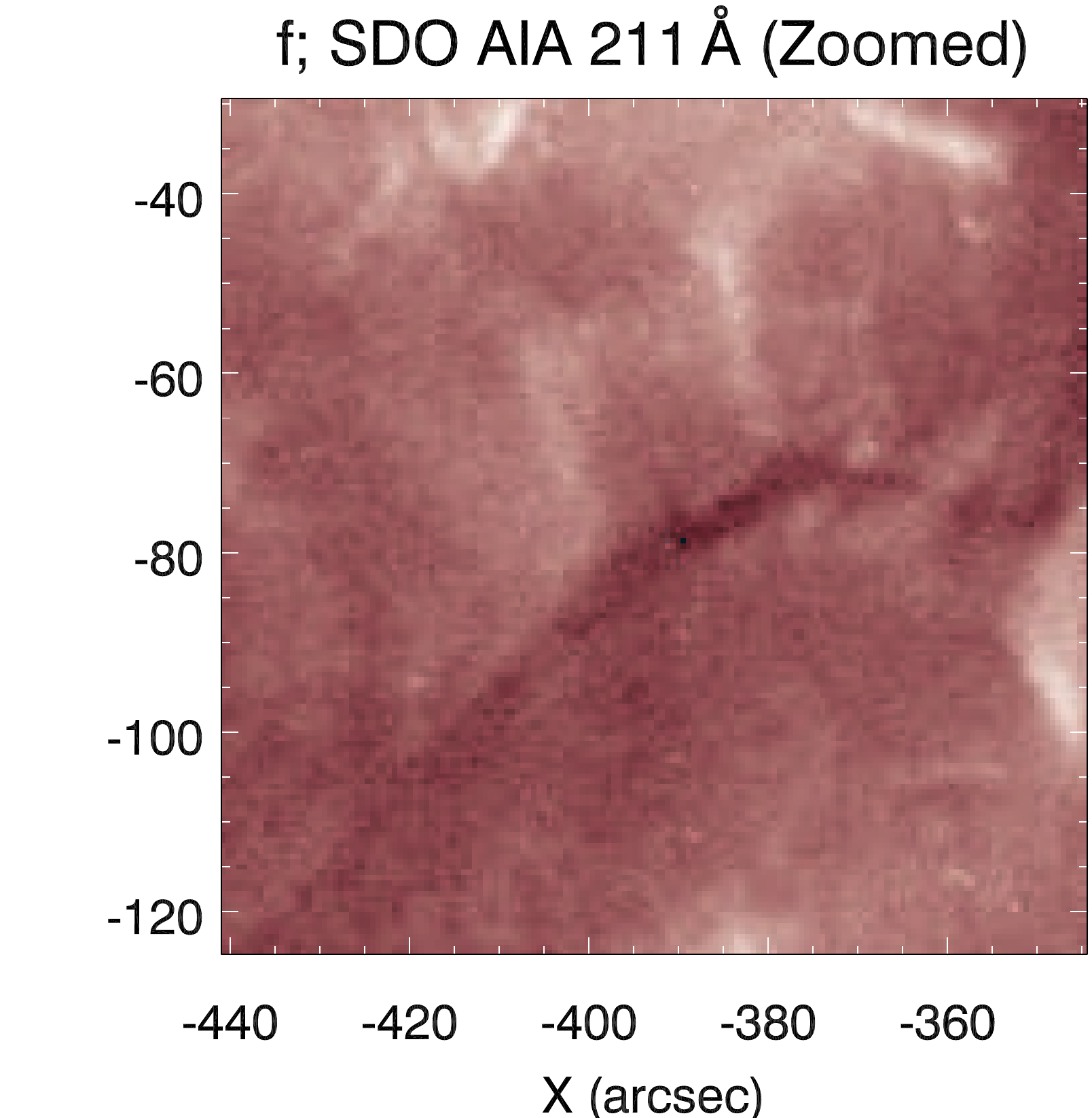}}}\ 
\caption{Filament observed on May 29, 2017. Panel \texttt{a}; 
context full\discretionary{-}{-}{-}disk image in the SDO/AIA 193\,\AA\ channel.
Panel \texttt{b}; zoom\discretionary{-}{-}{-}in on the box
marked in the panel \texttt{a}. Panel \texttt{c}; The H$\alpha$
line core image of the filament taken by DST/IBIS at 14:44:55\,UT.
White box indicates the region from which the average
quiet\discretionary{-}{-}{-}Sun profile was calculated.
Panels \texttt{d}\,--\,\texttt{f}; contextual SDO/AIA images in three EUV channels: 171\,\AA, 304\,\AA, and 211\,\AA.
The images have the same zoomed\discretionary{-}{-}{-}in field of view as is in panel \texttt{b}.}
\label{fig:context}
\end{figure*}
Solar filaments, or prominences as they are referred to
when observed projected above the solar limb, are an integral part of the
higher solar atmosphere. The filament plasma is believed to be embedded in the
coronal magnetic field in regions where the field is mostly horizontal 
and dipped. This type of magnetic field then supports the dense and cool filament
plasma against gravity and insulates it from the hot coronal
environment. Comprehensive reviews of the physics of prominences and filaments
can be found in \citet{2010SSRv..151..243L}, \citet{2010SSRv..151..333M}, in
the proceedings of the IAUS300 \citep{2014IAUS..300.....S}, in the review by
\citet{2018LRSP...15....7G}, or in the book 'Solar prominences' edited
by \citet{2015ASSL..415.....V}, for example. Filaments located within solar active regions are
usually short\discretionary{-}{-}{-}lived while quiescent
filaments located within the quiet\discretionary{-}{-}{-}Sun tend to have significantly 
longer life\discretionary{-}{-}{-}times of up to a few months.
Quiescent filaments are commonly observed in relation to filament channels 
\citep[see e.g.][]{2001ApJ...561L.223H,2003A&A...401..361S,2004SoPh..221..297S,2004A&A...421..323S}.
These filament channels are best viewed using spectral lines in Extreme Ultraviolet (EUV) part of the spectrum
and are able to out\discretionary{-}{-}{-}live their filament counterparts. This occurs because even though 
filament counterparts are stable for extended periods of time,
quiescent filaments can become unstable and erupt. Consequently, they frequently form
the core of coronal mass ejections \citep[CMEs, see e.g.][]{2011LRSP....8....1C}. As a
result, filament channels are often host to a sequence of consecutive filaments, which are most
readily visible in the H$\alpha$ line.  

It is generally understood that the driving force of the early stage of solar eruptions   
(moments after the global loss of equilibrium) is due to the magnetic field through mechanisms, such as
the torus instability, that is possibly triggered by the kink instability, and subsequently facilitated by
the reconnection \citep{Kliem:2006, Cheng:2017}. However, 
recent studies \citep[e.g.][]{Jenkins:2018, Jenkins:2019} highlight the ability of plasma within
pre\discretionary{-}{-}{-}eruptive filaments to exert some influence on the stability of the host magnetic field 
\citep[see also][]{2007ApJ...665..830P, 2011A&A...532A..93B, 2018ApJ...862...54F}. In particular, 
\citet{Jenkins:2019} underline the importance of draining the filament mass just prior to 
the global loss of equilibrium. It is therefore imperative to study the evolution of plasma properties within 
solar filaments, particularly those located within the quiet Sun (QS), in the lead up to their destabilisation,  
and subsequent eruption. However, in order to accurately quantify the temporal evolution of the plasma within 
these filaments, robust methods of deducing plasma properties in individual snapshots are first required. 

In this paper we study a small part of a larger quiescent filament (hereafter referred to as the observed
filament fragment), which remained stable for several days prior to its eruption. The
present work focuses on the quiescent stage of this filament where we are able to assume that the filament properties
do not significantly vary over timescales of hours or even a few days. For the interpretation of the H$\alpha$ spectra 
obtained at the observed fragment of the filament, we used a 2D non\discretionary{-}{-}{-}LTE radiative transfer model
(non\discretionary{-}{-}{-}LTE stands for departures from the LTE -- Local Thermodynamical Equilibrium). This model is based
on the 1D model developed by \citet{1998SoPh..179...75A,1999A&A...349..974A} for the modelling of
prominences. The prominence model was first generalised into 2D by 
\citet{2001A&A...375.1082H} and then used for the modelling of prominence fine structures by  
\citet{2005A&A...442..331H} and \citet{2008A&A...490..307G,2010A&A...514A..43G,2012A&A...543A..93G}.
The 1D non\discretionary{-}{-}{-}LTE radiative transfer model adapted for filaments was 
previously used by \citet{1999A&A...345..618M} for constructing the 2D maps of  
physical properties of a filament observed spectroscopically in the H$\alpha$ line. 
Later, a similar technique was applied by \citet{2001A&A...366..686T} for the analysis of chromospheric 
cloud\discretionary{-}{-}{-}like structures (including filaments) observed in the \ion{Ca}{ii} 8542\,\AA\ line. 
More recently, \citet{2006A&A...459..651S,2012SoPh..281..707S} used the 1D non\discretionary{-}{-}{-}LTE model 
to analyse filaments observed in the hydrogen Lyman series and H$\alpha$ line. 
The 2D prominence model of \citet{2001A&A...375.1082H} was generalised for adoption of the filament geometry and
used by \citet{2016AN....337.1045S} to analyse H$\alpha$ observations of an active region filament. These authors introduced
two significant ($25\,\mathrm{km/s}$) opposing line\discretionary{-}{-}{-}of\discretionary{-}{-}{-}sight (LOS) velocities, 
which were needed to reproduce the observed, very broad and symmetric profiles. 

In the observed filament fragment studied in the present work, the occurrence of significantly
asymmetric profiles in the observed H$\alpha$ data supports the use of a single LOS\discretionary{-}{-}{-}velocity value
as an input parameter of the model. This approach can be applied as it was already shown by 
\citet{1998Natur.396..440Z} that asymmetric H$\alpha$ profiles in prominences are caused by systematic 
plasma flows. Unresolved motions of the filament plasma can then be represented
by the micro\discretionary{-}{-}{-}turbulent velocity as another input parameter of the model.

The paper is organised as follows. In Sect.~\ref{s:observations} we describe the observations of the
studied filament fragment together with the processes of data calibration and
co\discretionary{-}{-}{-}alignment. In Sect.~\ref{s:nonltemodel} we introduce the 2D
non\discretionary{-}{-}{-}LTE filament model that produces the synthetic H$\alpha$
spectra, which we compare with the observations. The results
obtained by the modelling are presented in Sect.~\ref{s:modelresults} and are
subsequently discussed in Sect.~\ref{s:discussion}. In this section we also discuss the
influence of background radiation on the derived parameters of
the observed filament plasma and we assess the uncertainties of
these parameters. In Sect.~\ref{s:Conclusions} we offer our
conclusions.
%
%
\section{Observations}\label{s:observations}
%
\subsection{General information on observations}
\label{s:obs_gen_info}
A fragment of a larger solar filament was observed on May 29, 2017. 
The observed filament fragment was located at the heliographic position around $-24.3\,\deg$, $-4.85\,\deg$ 
(i.e. solar~X\,=\,$-400\,\mathrm{arcsec}$, solar~Y\,=\,$-80\,\mathrm{arcsec}$).
This specific fragment was chosen because it was a part of the only
filament on\discretionary{-}{-}{-}disk on the day and its strong absorption signature
and interesting shape positioned at the end of the filament made for an interesting target. 
The filament was initially quiescent but became
activated and erupted on May 30, 2017.
Exact time of the eruption cannot be stated as its progress was rather slow: According
to observations taken in EUV by the EUV Imager \citep[EUVI; ][]{2004SPIE.5171..111W} 
instrument on board the Solar Terrestrial Relations Observatory\discretionary{-}{-}{-}A    
(STEREO\discretionary{-}{-}{-}A) satellite in the $304$\,\AA\ channel,
the filament seen as prominence starts to dramatically
increase in height from between 10:00 and 12:00\,UT. 
The ongoing eruption was also observed during the DST campaign from 13:47\,UT and by the 
Atmospheric Imaging Assembly \citep[AIA;][]{2012SoPh..275...17L} instrument on board the Solar
Dynamics Observatory \citep[SDO;][]{2012SoPh..275....3P} in its all EUV channels.  
We want to emphasise here that in the present paper we focus on the quiescent stage of its evolution
as observed on May 29, 2017. A study of the erupting phase of this filament will be a subject of 
future work. 

Figure~\ref{fig:context}\texttt{a} shows a full\discretionary{-}{-}{-}disk
context image of the observed filament taken in the EUV at 193~\AA\ by the AIA instrument. 
Figure~\ref{fig:context}\texttt{b} shows a close\discretionary{-}{-}{-}up view in the same
channel of the AIA instrument and Fig.~\ref{fig:context}\texttt{c} shows the filament in
the H$\alpha$ line core observed at 14:44:55\,UT using the
Interferometric Bidimensional Spectropolarimeter
\citep[IBIS;][]{2006SoPh..236..415C} mounted at  the Dunn Solar Telescope (DST).
IBIS is a dual Fabry\discretionary{-}{-}{-}P\'erot imaging spectrometer capable
of taking narrow\discretionary{-}{-}{-}band images in the range of
$5,800$\,--\,$8,600$\,\AA\ alongside broad\discretionary{-}{-}{-}band
images that are useful for context and alignment purposes. With a
$95$\,arcsec diameter field of view (FOV) illuminating a $1,024\,\times\,1,024$
square\discretionary{-}{-}{-}pixel CCD, IBIS is diffraction limited across a 
sub\discretionary{-}{-}{-}window of $1,007\,\times\,1,007$\,pixels, 
resulting in a maximum spatial sampling of
$\approx\,0.1\,\mathrm{arcsec}\times\,0.1\,\mathrm{arcsec}$ per pixel.
However, due to the atmospheric seeing, the actual spatial resolution of the IBIS observation must
be measured by analysing the 2D Fourier power of the IBIS line\discretionary{-}{-}{-}core intensity image as a function
of the spatial resolution. It was found that the Fourier power decreased with finer resolution until a value of
$0.7$\,arcsec where the Fourier power became more\discretionary{-}{-}{-}or\discretionary{-}{-}{-}less constant (except
for some noise). Thus, this boundary value of resolution indicates the smallest scale resolvable in
the line\discretionary{-}{-}{-}core image of the instrument, which is the spatial resolution of the observation. 
Equipped with a high\discretionary{-}{-}{-}order adaptive optics system
\citep[see][for a detailed description]{2004ApJ...604..906R},
the DST allows IBIS to capture seeing\discretionary{-}{-}{-}corrected
images of the Sun with high temporal cadence. Panels \texttt{d}\,--\,\texttt{f} of the figure show 
the studied filament fragment in additional three AIA EUV channels -- 171, 304, and 211\,\AA\ -- zoomed
in the same FOV as in the panel \texttt{b}. 

The filament was observed at the DST for several days prior to its eruption, which occurred on the day following the
snapshot shown here. Observations of the IBIS instrument made during May 29, 2017 were carried out in
three blocks at the following times: 13:59:46\,--\,15:30:26, 15:30:52\,--\,15:42:12, and 20:59:00\,--\,22:18:10\,UT. 
The snapshot shown in Fig.~\ref{fig:context}c was taken at 14:44:55\,UT and represents the best seeing available
within the obtained dataset. Therefore the IBIS data from this time were used in the filament modelling. 
The FOV of IBIS did not cover the entire filament which was approximately 660\,arcsec 
long. The FOV in Figs.~\ref{fig:context}\texttt{b} and \texttt{c} contains 
what we believe to be one of the footpoints of the filament and
its surroundings. The co\discretionary{-}{-}{-}alignment
of the H$\alpha$ line core image (Fig.~\ref{fig:context}c) with
the SDO/AIA 193\,\AA\ image (Fig.~\ref{fig:context}b) is only approximate.

\begin{figure}
\resizebox{0.5\textwidth}{!}{
\includegraphics{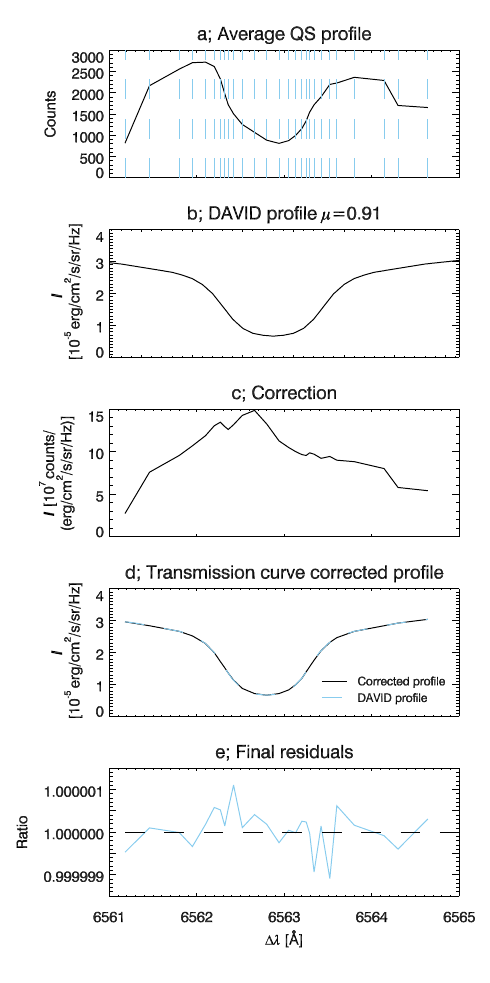}}
\caption{Illustration of the pre\discretionary{-}{-}{-}filter
correction method using the H$\alpha$ tables of
\citet{1961ZA.....53...37D}. Panel \texttt{a}; Average QS profile
calculated from the area marked in
Fig.~\ref{fig:context}\texttt{c}. Vertical dashed lines indicate 27 IBIS
wavelength positions used. Panel \texttt{b}; Reference David's
H$\alpha$ profile at $\mu_{\mathrm{cal}}=0.91$. Panel \texttt{c}; Correction
between panels \texttt{a} and \texttt{b}. Panel \texttt{d};
Average QS profile with panel \texttt{c} correction applied, and
panel \texttt{b} over\discretionary{-}{-}{-}plotted in
dashed\discretionary{-}{-}{-}line. Panel \texttt{e}; Residuals
remaining between the corrected profile and the reference
profile.}
\label{fig:pre-method}
\end{figure}

In this paper, we use a single data segment
consisting of observations of both H$\alpha$ ($6562.8$\,\AA) and
\ion{Ca}{ii} ($8542$\,\AA) lines. For H$\alpha$, 27 wavelength
positions were selected between 6561.18 and 6564.64~\AA\ such that 
there was a higher density of chosen wavelength positions 
in the near wings of the profile than in the farther wings and 
in the core. This can be seen in Fig.~\ref{fig:pre-method}a. 
Another 30 wavelength positions were used to scan the \ion{Ca}{ii} 
line. With an exposure time for each wavelength position of 80\,ms, 
a data segment (one full scan of both H$\alpha$ and \ion{Ca}{ii}) was
completed every $\approx\,13$\,s. This work is devoted only to
diagnostics of physical properties of the hydrogen plasma of the observed
filament, thus, non\discretionary{-}{-}{-}LTE modelling of the \ion{Ca}{ii}
line is outside of the scope of this paper. A similar 2D slab filament model
for the \ion{Ca}{ii} ion is now in preparation and is planned as the
focus of our future papers. 
%
%
\begin{figure}
\resizebox{0.47\textwidth}{!}{
\includegraphics{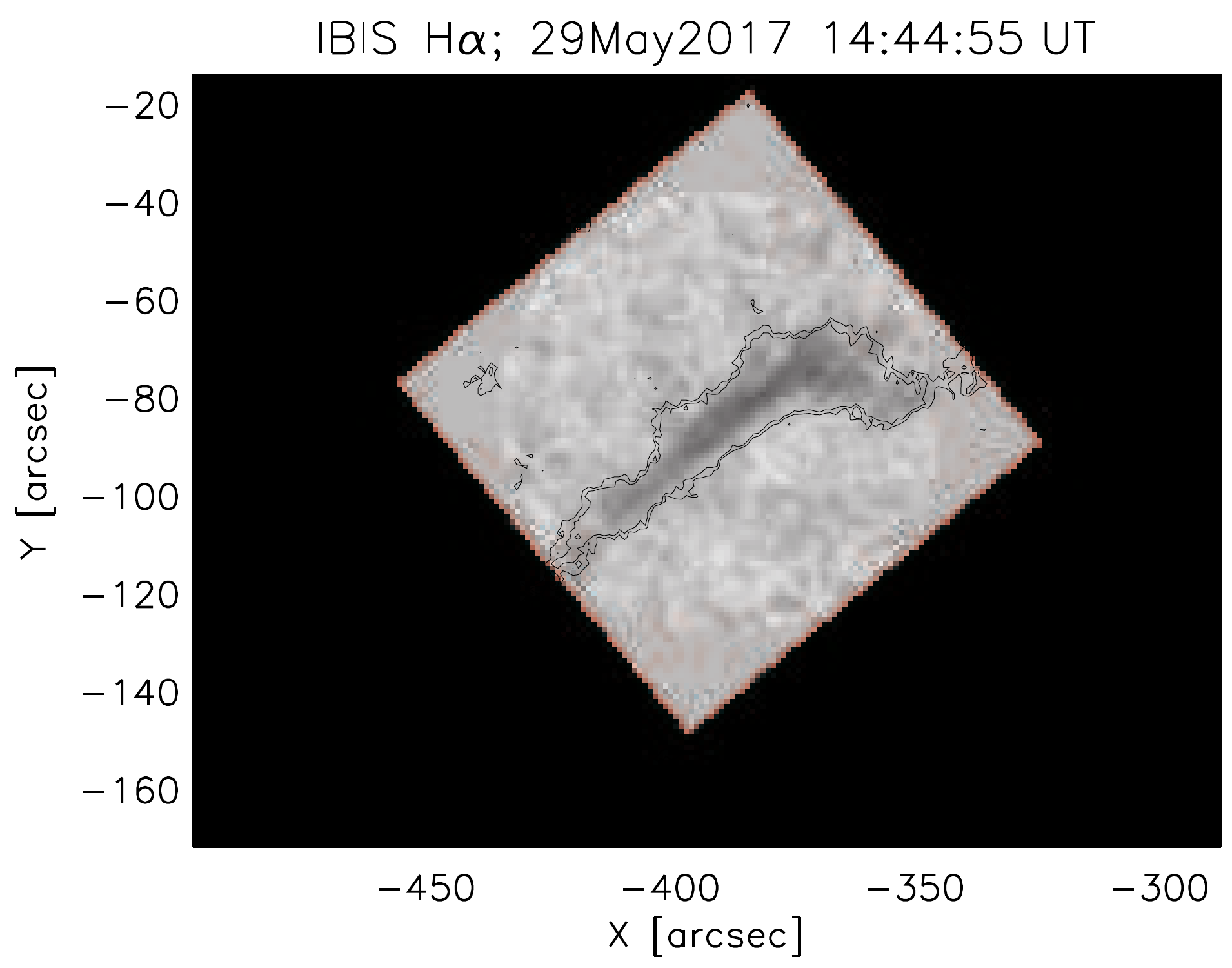}}\\ 
\resizebox{0.47\textwidth}{!}{
\includegraphics{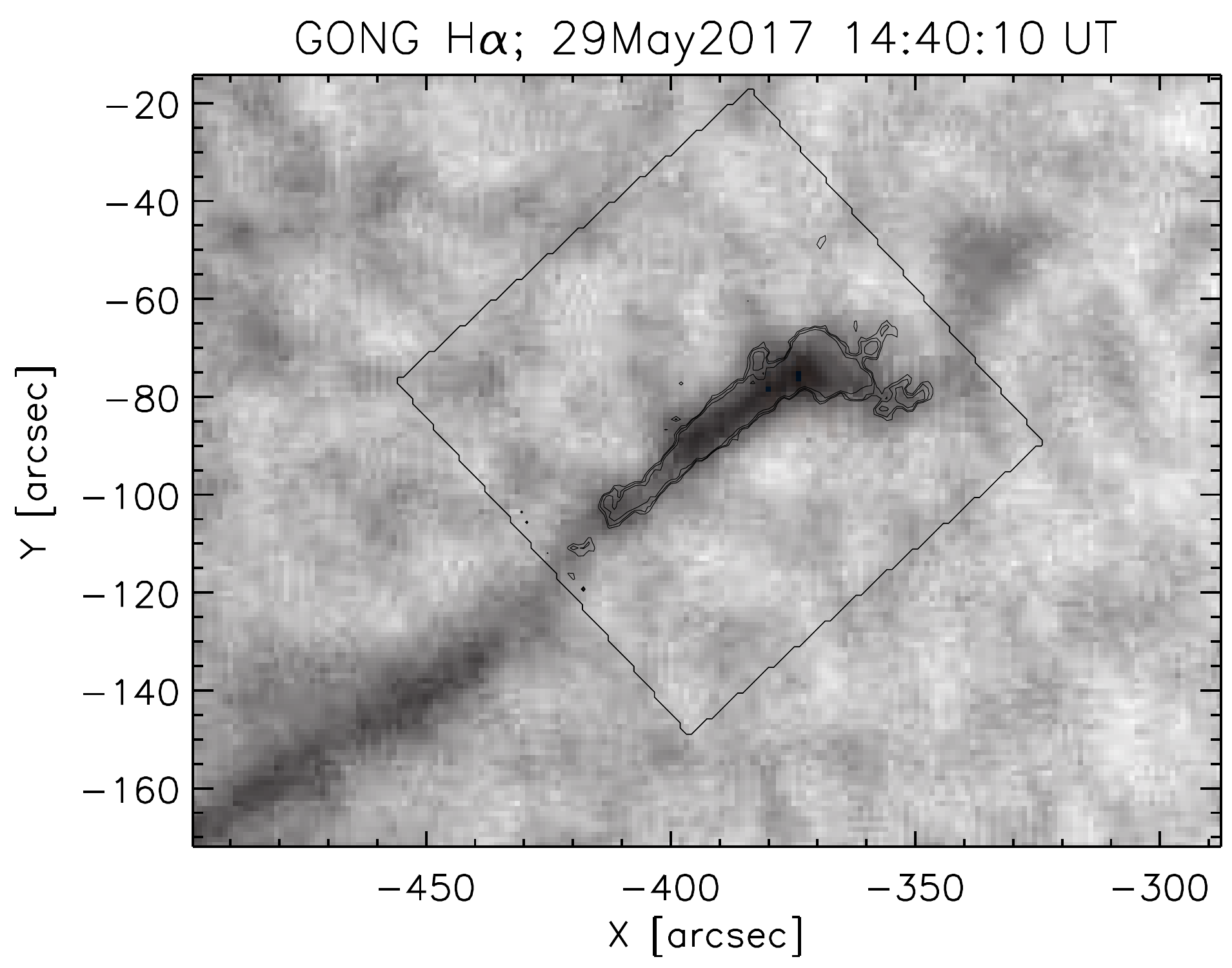}} 
\caption{Co\discretionary{-}{-}{-}alignment of the IBIS H$\alpha$ data 
(\textit{top}\/) with the full\discretionary{-}{-}{-}disk H$\alpha$ observations
obtained by GONG (\textit{bottom}\/). The black
square in the bottom panel marks the FOV of the IBIS observations.
Contours plotted in the GONG image outline the IBIS intensity map and
vice versa. Any difference in contrast at the filament as seen 
in these two images are caused by different transmission of both 
instruments.}
\label{fig:coal}
\end{figure}
%
\subsection{Data reduction and co-alignment}
\label{s:datacalib}
The narrow\discretionary{-}{-}{-}band (NB) and
broad\discretionary{-}{-}{-}band (BB) data obtained by IBIS
were processed using the IBIS data processing
pipeline\footnote{\url{https://pdfs.semanticscholar.org/d480/614334a008e35765f935b397147b12bd679c.pdf}}.
The raw data were first corrected for the dark currents induced by
the electronics associated with the detector. Gain tables were
then calculated from flat field observations to characterise the
sensitivity of the detector across its face. Detector image scale
and relative rotation between NB and BB channels were determined
from dot\discretionary{-}{-}{-}line grid images taken during the
observing run. Finally, the systematic wavelength shift induced
by the collimated mounting of the Fabry-P\'erot interferometer
was removed from the NB data.
%
%
\begin{figure*}
\centering
\resizebox{0.9\textwidth}{!}{
\includegraphics{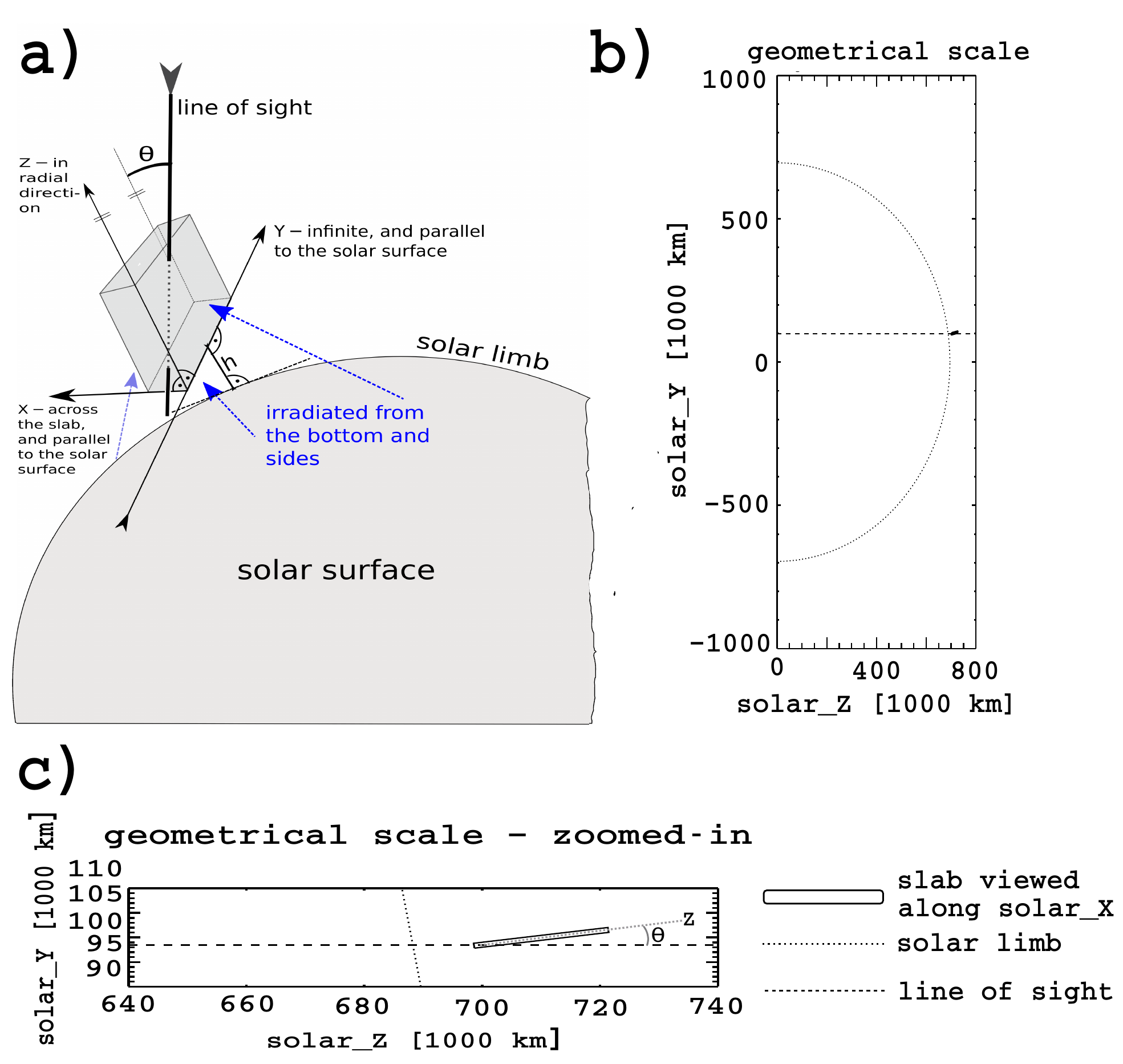}} 
\caption{Geometrical scheme of the non\discretionary{-}{-}{-}LTE
2D\discretionary{-}{-}{-}slab filament model and geometry of the
solar\_Y\discretionary{-}{-}{-}solar\_Z cross\discretionary{-}{-}{-}section 
of the 2D slab in a large scale encompassing nearly half of the Sun shown
in panels \texttt{a} and \texttt{b}, respectively. 
In order to distinguish the coordinate system of the slab itself from
the helioprojective coordinate system (Cartesian coordinate system with origin
in the Sun centre and its X\discretionary{-}{-}{-}axis pointing to the solar W), 
the first one was denoted by simple X, Y, and Z letters while for
the latter these three letters are preceded by the 'solar\_' prefix. 
The panel \texttt{c} shows the same geometrical scheme in more
detail. While just a schematic cartoon explaining the
2D slab\discretionary{-}{-}{-}model geometry is presented in
the panel \texttt{a}, schemes in the panels \texttt{b} and
\texttt{c} are produced as a graphical output of the model
code. Size of the $\theta$ angle is exaggerated in all three
panels for its better recognition, in reality $\theta$ is much smaller
for the studied filament fragment. 
} \label{fig:model_geom_scheme}
\end{figure*}
A comparison of the observed H$\alpha$ profiles with synthetic
H$\alpha$ profiles provided by non\discretionary{-}{-}{-}LTE
modelling requires the observed data to be radiometrically
calibrated into the physical units. To do so, the correction for
the influence of the pre\discretionary{-}{-}{-}filter transmission
curve on the shape of the NB line profiles needs to be taken into
account. The calibration coefficient was derived by a comparison
in a filament close\discretionary{-}{-}{-}by QS area 
with the H$\alpha$ reference profile constructed from the table of
\citet{1961ZA.....53...37D} for the angle between normal to the
solar surface at the QS area position and the LOS of 
$\mu_{\mathrm{cal}}$=$0.91$, which was measured as a length of
the great arc between this position and disk centre. The QS
profile (Fig.~\ref{fig:pre-method}\texttt{a}) was obtained as an
average from a quiet region next to the filament which is marked
by a white box in Fig.~\ref{fig:context}c. 
Dispersion of spectral intensities of individual
profiles observed inside the quiet region is only up to $\pm40$\,\% of the QS profile
intensities. Almost 80\,\% of profiles from the quiet region are shifted in wavelength
by less than $0.1\,\mathrm{km/s}$ and the remaining profiles are shifted by less than
$7\,\mathrm{km/s}$. Thus, selection of the quiet region is reliable for the QS profile
as it was averaged mostly from similar profiles not influenced much by velocities. 
The reference H$\alpha$ profile is shown in Fig.~\ref{fig:pre-method}\texttt{b}.
The correction for the pre\discretionary{-}{-}{-}filter transmission profile takes
into account the dependence of the calibration coefficient on the
wavelength. The calibration coefficient as a function of
wavelength (see Fig.~\ref{fig:pre-method}\texttt{c}) is applied for the
calibration\discretionary{-}{-}{-}correction of all observed IBIS
H$\alpha$ profiles. Panels \texttt{d} and \texttt{e} of
Fig.~\ref{fig:pre-method} show how well the QS profile fits the
reference H$\alpha$ profile after applying the
calibration\discretionary{-}{-}{-}correction process. The relative 
errors $\sigma_{\mathrm{rel}}(\lambda)$ of the specific intensities
$I(\lambda)$ calibrated to $\mathrm{erg/cm}^2\mathrm{/s/sr/Hz}$ were 
calculated using the Poisson statistics:
\begin{equation} 
\sigma_{\mathrm{rel}}(\lambda)=\frac{\sqrt{I_{\mathrm{cnts}}(\lambda)\rule{0pt}{1.6ex}}}{I_{\mathrm{cnts}}(\lambda)}
\label{eq:calibinterreq}
\end{equation}
where $I_{\mathrm{cnts}}(\lambda)$ are the non\discretionary{-}{-}{-}calibrated specific intensities in counts.

%
%
\begin{figure*}
\centering
\resizebox{0.97\textwidth}{!}{
\includegraphics{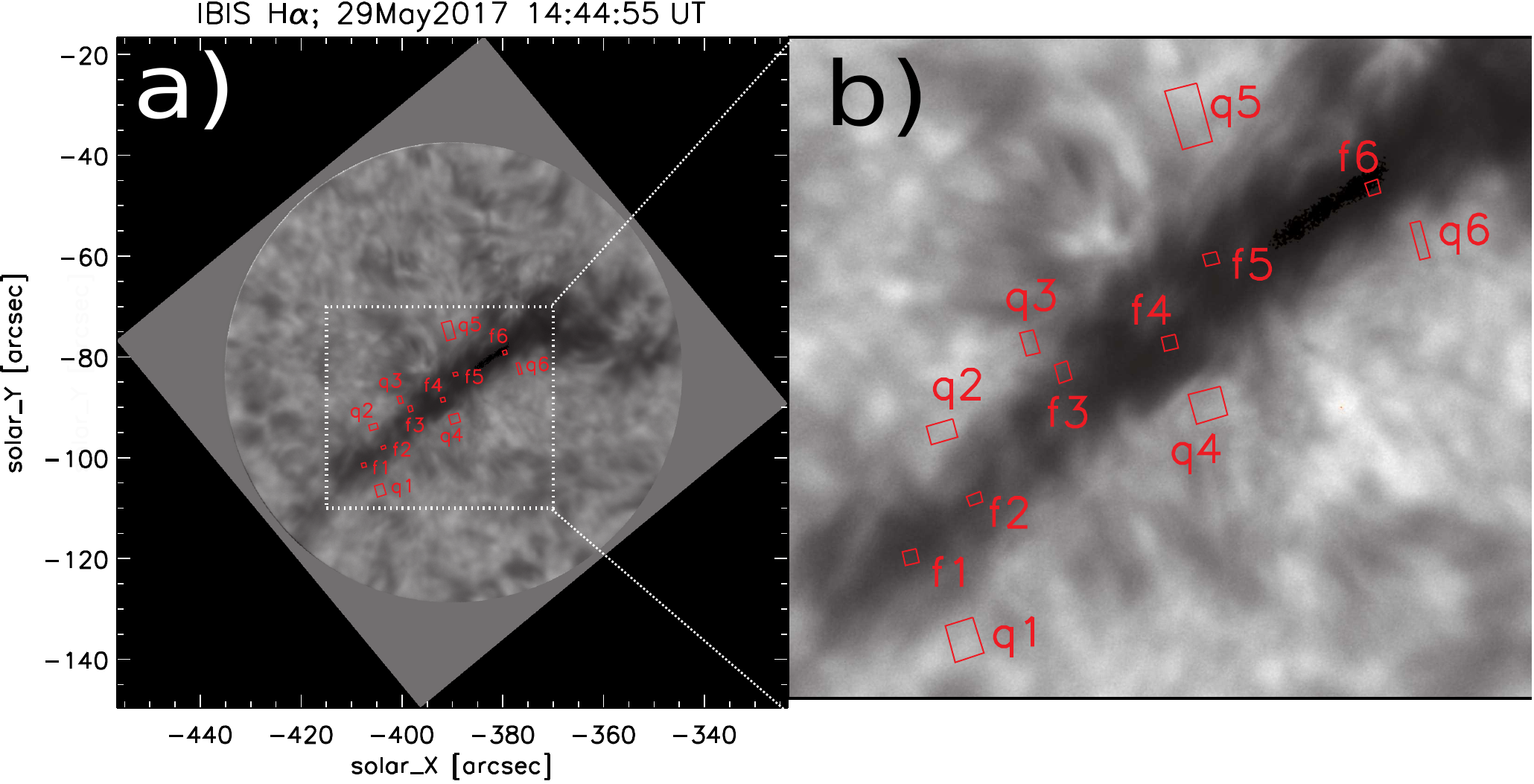}}
\caption{Maps of integrated intensities of the H$\alpha$ line core
(within $\Delta\lambda$=$\pm0.40$\,\AA) obtained by IBIS.
Marked are the six areas f1\,--\,f6 within the observed filament fragment and the six
associated QS areas q1\,--\,q6 in their vicinity. The whole FOV of IBIS is shown in
panel \texttt{a} and the zoomed area around the 
observed filament fragment is shown in panel \texttt{b}.} \label{fig:sixareasimap}
\end{figure*}
To correct the orientation and pointing of the IBIS observations
we co\discretionary{-}{-}{-}aligned them with the
full\discretionary{-}{-}{-}disk H$\alpha$ data obtained by  
instruments of the Global High Resolution H\discretionary{-}{-}{-}alpha 
Network belonging to the Global Oscillation Network Group
\citep[GONG;][]{2011SPD....42.1745H} of the National Solar
Observatory (NSO). To do so, we used a full\discretionary{-}{-}{-}disk map of
intensities integrated within wavelength range
$\Delta\lambda$=$\pm0.4\,\AA$ around the H$\alpha$ line centre.
The IBIS data were vertically flipped, rotated by $50.4\,\deg$ in
clock\discretionary{-}{-}{-}wise direction and the centre of the
FOV was positioned at $-389.9\,\mathrm{arcsec}$ in solar~X and
$-83.0\,\mathrm{arcsec}$ in solar~Y. The comparison of the
co\discretionary{-}{-}{-}aligned IBIS H$\alpha$ intensity map with
the corresponding cut\discretionary{-}{-}{-}off from the
full\discretionary{-}{-}{-}disk GONG H$\alpha$ image is shown in
Fig.~\ref{fig:coal}. To demonstrate the reliability of the
co\discretionary{-}{-}{-}alignment, contours of the filament in
the GONG H$\alpha$ image are plotted on the IBIS H$\alpha$
intensity map and vice versa.
%
%
%
\section{2D non-LTE filament model}\label{s:nonltemodel}
At this point we should comment on the assumed 2D geometry for our filament study. Using 1D horizontal
slab models, \citet{1999A&A...345..618M} and \citet{2001A&A...366..686T} constructed 2D maps of plasma 
parameters in filaments observed in the H$\alpha$ and \ion{Ca}{ii} 8542 \AA\ lines, respectively. The
line of sight was intersecting a horizontally\discretionary{-}{-}{-}infinite 1D slabs where the vertical
extension was a free parameter.
MALI1D code was used to compute a large grid of isothermal\discretionary{-}{-}{-}isobaric models
and then the filament parameters were derived by the least\discretionary{-}{-}{-}square fitting of the observed
profiles. After the grid is constructed, this method is very fast and one can make maps of parameters
pixel\discretionary{-}{-}{-}by\discretionary{-}{-}{-}pixel. However, a disadvantage is that the 1D horizontal
slab was illuminated from the solar disk only at its bottom, while in real prominences the incident radiation
penetrates through filamentary structures which are then illuminated from all directions (not from corona, where
H$\alpha$ and \ion{Ca}{ii} radiation in not generated). Although a 2D slab model of the whole filament was
introduced by \citet{1993A&A...274..571P} and \citet{1995A&A...302..587P} (and even earlier by
\citet{1982ApJ...254..780V}\,), these authors did not analyse 2D high\discretionary{-}{-}{-}resolution spectra.
Looking at filaments at very high resolution (e.g. in H$\alpha$), we see many fine\discretionary{-}{-}{-}structure
threads aligned at some angle to the filament axis. These features were modelled by \citet{2006ApJ...643L..65H}
as 2.5D magnetic dips seen against the solar disk as a filament.
Their approach allowed only the lateral illumination of vertically infinite fine\discretionary{-}{-}{-}structure
threads to be considered. 
In this paper we use another 2D geometry and represent a local fine structure of the filament
(i.e. at a given position of observation) by a thin ($1000$\,km) 2D horizontal slab
(see Fig.~\ref{fig:model_geom_scheme}). 
This allows illumination both from the bottom and from both sides of the 2D slab, critical for the determination
of the line source functions. However, we neglect possible effect of a mutual irradiation between fine structures,
such effects have not been studied in case of prominences.

The 2D non\discretionary{-}{-}{-}LTE filament model used in this
work is based on the 2D radiative transfer \texttt{MALI2D} code developed by
\citet{2001A&A...375.1082H} for modelling of prominences. This code 
was modified here to represent the geometry of filaments. The
radiative transfer is solved using the
short\discretionary{-}{-}{-}characteristics method
\citep{1988JQSRT..39...67K} together with the Multilevel
Accelerated Lambda Iterations
\citep[MALI;][]{1991A&A...245..171R}. The statistical equilibrium is
calculated for a 5\discretionary{-}{-}{-}level plus continuum
hydrogen atom. The partial frequency redistribution
\citep[see e.g.][]{1987A&A...183..351H} is used in calculations of
the Ly$\alpha$ and Ly$\beta$ lines while the complete frequency
redistribution is applied for all other spectral lines of
hydrogen. For the formal solution of the radiative transfer along
a line of sight we used the method of \citet{1978ApJ...220.1001M}
and average H$\alpha$ profiles from QS areas located in the
close vicinity of the observed filament are used as the background
radiation.

The filament was approximated by an isothermal and isobaric 2D slab
placed horizontally above the solar surface. The infinite
horizontal dimension was along the Y\discretionary{-}{-}{-}axis and the finite dimensions
(X) and (Z) formed the cross\discretionary{-}{-}{-}section of the filament (see
Fig.~\ref{fig:model_geom_scheme}). The vertical, more extended dimension Z of
the slab was oriented in the radial direction. 
The 2D slab was irradiated from the solar surface at the bottom and
the sides. The dilution effect was properly taken into account.
For irradiation of the slab at its bottom edge, dilution for the bottom
height above the solar surface of the slab was calculated taking into account limb
darkening for the spectral lines H$\alpha$\,--\,H$\gamma$, Paschen~$\alpha$ and $\beta$, and
Bracket~$\alpha$. For hydrogen resonance (Lyman) lines plus continuum the limb darkening
was assumed to be negligible. For irradiation from sides, irradiation from a half\discretionary{-}{-}{-}plane
(only half of the solar disc is seen from the slab side) was taken, and individual values of dilution factors
were calculated for different heights along a side of the slab. The effect of the limb
darkening was also taken into account for the same spectral lines as for irradiation from the bottom.  
The following QS data were used for the irradiation of the slab: the
Balmer series line profiles for different values of $\mu$ from \citet{1961ZA.....53...37D}; Ly$\alpha$ line
profiles observed by the Laboratoire de Physique Stellaire et Planetaire (LPSP) instrument \citep{1978ApJ...221.1032B}
on board the Orbiting Solar Observatory 8 (OSO\discretionary{-}{-}{-}8) satellite from
\citet{1978ApJ...225..655G, 1981A&A...103..160L}; the higher Lyman lines from \citet{1998ApJS..119..105W} 
observed by the Solar Ultraviolet Measurements of Emitted Radiation (SUMER) spectrograph \citep{1995SoPh..162..189W}
on board the SOlar and Heliospheric Observatory (SoHO) satellite.

The radiation emergent from the model was computed for a LOS 
direction which was inclined from the radially oriented
Z\discretionary{-}{-}{-}axis of the slab by an angle
$\theta$, which was derived from the position of the observed
filament on the solar disk as shown in the geometrical scheme
in panels \texttt{b} and \texttt{c} of Fig.~\ref{fig:model_geom_scheme}.
The slab is thus seen in a projection onto the solar disk. The position
along the projection of the slab is expressed in km from its south end.
Size of the angle $\theta$ in the scheme is exaggerated for
better understanding. The cosine of the angle $\theta$ is
hereafter referred to as $\mu$. In reality the angle $\theta$ 
is only around $0.20\,\deg$ ($\mu$ almost equal to unity) 
at the filament position.
As it is shown further, the best fit to observed data was obtained
for a vertical side of the 2D slab projected under this very small angle, 
not for the horizontal area on top of the 2D slab. It is necessary to note 
that this angle is different from $\mu_{\mathrm{cal}}$ of $0.91$ that referred
previously to the position of the QS reference profile used for absolute
calibration (see Sect.~\ref{s:datacalib}).
%
\section{Results obtained by modelling}\label{s:modelresults}
%
%
%
\begin{table*}
\centering{
\caption{Input parameters (and their uncertainties) of 2D
models providing the best fit to the observed H$\alpha$
profiles from the areas
f1\,--\,f6: vertical dimension of the slab (the slab
Z\discretionary{-}{-}{-}axis), temperature, gas pressure,
microturbulence velocity and the LOS component of the
plasma\discretionary{-}{-}{-}flow velocity. Positive
value of the LOS velocity corresponds to flow towards an observer.}
\label{table:resmodparams}
\begin{tabular}{ c c c | r r r r r }
\hline \\[-2.8ex]
  & \multicolumn{2}{c|}{position} & \multicolumn{1}{l}{vertical size}      &                                 & \multicolumn{1}{l}{gas}                   &                                         & 
                                      \\[-0.5ex]
area & solar X & solar Y & \multicolumn{1}{l}{of the slab (Z)}    & \multicolumn{1}{c}{temperature} & \multicolumn{1}{l}{pressure}              & \multicolumn{1}{c}{$v_{\mathrm{MT}}$} &
  \multicolumn{1}{c}{$v_{\mathrm{LOS}}$}  \\
 & $\left[\mathrm{arcsec}\right]$ & $\left[\mathrm{arcsec}\right]$  & \multicolumn{1}{c}{[km]}               & \multicolumn{1}{c}{[K]}         & \multicolumn{1}{c}{$\left[\mathrm{dyn/cm}^2\right]$} & \multicolumn{1}{c}{$\left[\mathrm{km/s}\right]$} &
  \multicolumn{1}{c}{$\left[\mathrm{km/s}\right]$}           \\[0.7ex]
\hline \\[-2.8ex]
f1  & $-408$ & $-101$ & $33,000\pm12\,\%$ & $10,000\pm10\,\%$ & $0.28\pm43\,\%$         & $9.33\pm45\,\%$     & $-6.02\pm7\,\%\hspace{2.2ex}$  \\
f2  & $-404$ &  $\ -98$ & $38,000\pm10\,\%$ & $9,000\pm11\,\%$  & $0.20\pm50\,\%$         & $4.30\pm93\,\%$     & $-6.87\pm5\,\%\hspace{2.2ex}$  \\
f3  & $-398$ &  $\ -90$ & $37,500\pm60\,\%$ & $6,000\pm17\,\%$  & $0.42\pm67\,\%$         & $8.03\pm20\,\%$     & $-7.35\pm10\,\%\hspace{1.1ex}$ \\
f4  & $-392$ &  $\ -88$ & $22,500\pm15\,\%$ & $12,000\pm8\,\%\hspace{1.1ex}$  & $0.15\pm40\,\%$         & $6.30\pm82\,\%$     & $-1.31\pm83\,\%\hspace{1.1ex}$ \\
f5  & $-398$ &  $\ -83$ & $32,000\pm22\,\%$ & $11,000\pm9\,\%\hspace{1.1ex}$  & $0.16\pm56\,\%$         & $4.87\pm88\,\%$     & $-2.53\pm37\,\%\hspace{1.1ex}$ \\
f6  & $-379$ &  $\ -79$ & $37,500\pm4\,\%\hspace{1.1ex}$  & $14,000\pm7\,\%\hspace{1.1ex}$  &
$0.16\pm56\,\%$ & $4.21\pm43\,\%$ & $-0.10\pm502\,\%$    \\
\hline
\end{tabular}
}
\end{table*}
%
%
For the analysis, we chose six areas (f1\,--\,f6) located within 
the observed filament fragment (see Fig.~\ref{fig:sixareasimap}).
According to the general shape of the filament as seen in the H$\alpha$ full\discretionary{-}{-}{-}disc
images obtained by the GONG instrument (not shown here), one can assume that
in a hooked part of the observed filament fragment at the position in solar~X from approx. $-360$\,arcsec to $-375$\,arcsec
and in solar~Y from approx. $-65$\,arcsec to $-78$\,arcsec, a footpoint or barb could occur.
Assuming that the magnetic field
is more vertical in a footpoint, H$\alpha$ profiles from this part of the filament cannot be modelled accurately with
our non\discretionary{-}{-}{-}LTE model represented by a horizontal slab. 
Therefore the areas f1\,--\,f6 were chosen along the observed filament fragment outside the footpoint area.
No special approach was used for selection of positions for the six filament areas, they were just distributed approximately along
the axis of the filament fragment. Areas f1, f2, f4, and f6 were located more inside of the filament dark structure,
while f3 and f5 were positioned closer to its northern edge. The average H$\alpha$ profiles from these areas were fitted by the 2D
non\discretionary{-}{-}{-}LTE model described in the previous section.
We chose also six associated QS areas (q1\,--\,q6) in the vicinity
of the filament (see Fig.~\ref{fig:sixareasimap}) from which the average
profiles were used as the background radiation in the formal
radiative transfer solution.
%
\begin{figure*}
\parbox{\columnwidth}{
\resizebox{\columnwidth}{!}{
  \includegraphics{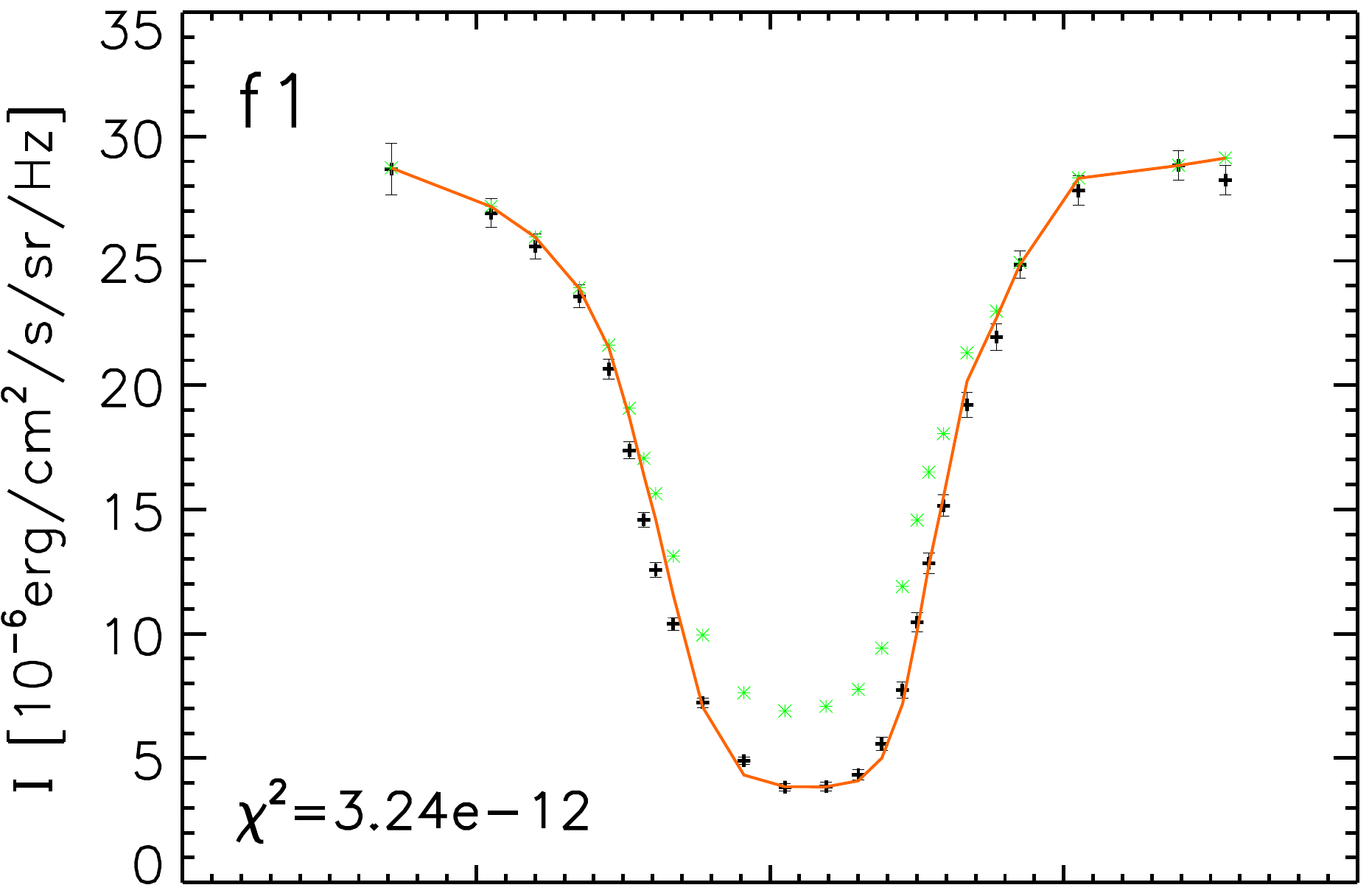}}}\ 
\parbox{0.11\columnwidth}{\phantom{1ex}}\
\parbox{0.89\columnwidth}{
\resizebox{0.89\columnwidth}{!}{
  \includegraphics{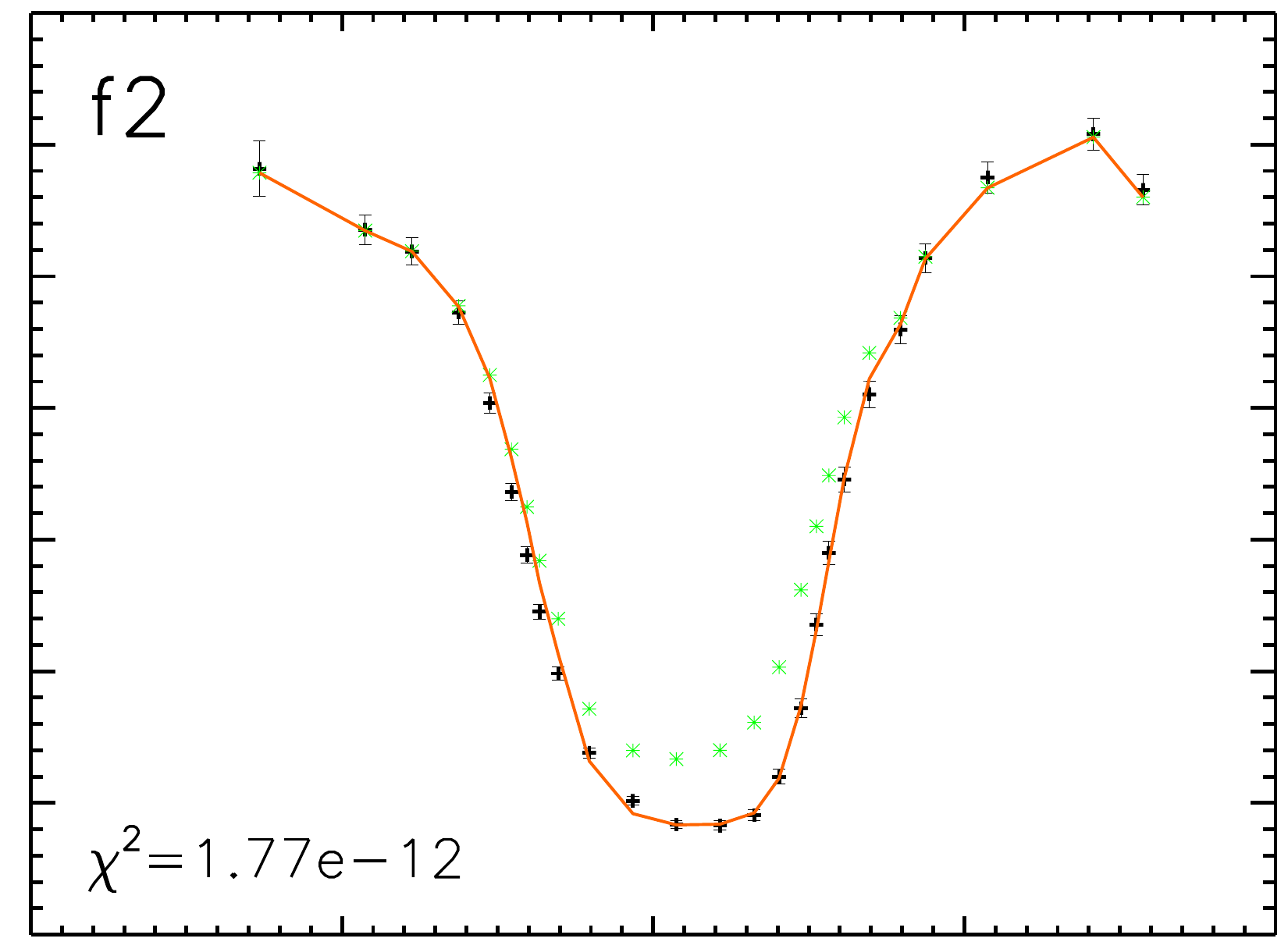}} 
}\\
\parbox{\columnwidth}{
\resizebox{\columnwidth}{!}{
  \includegraphics{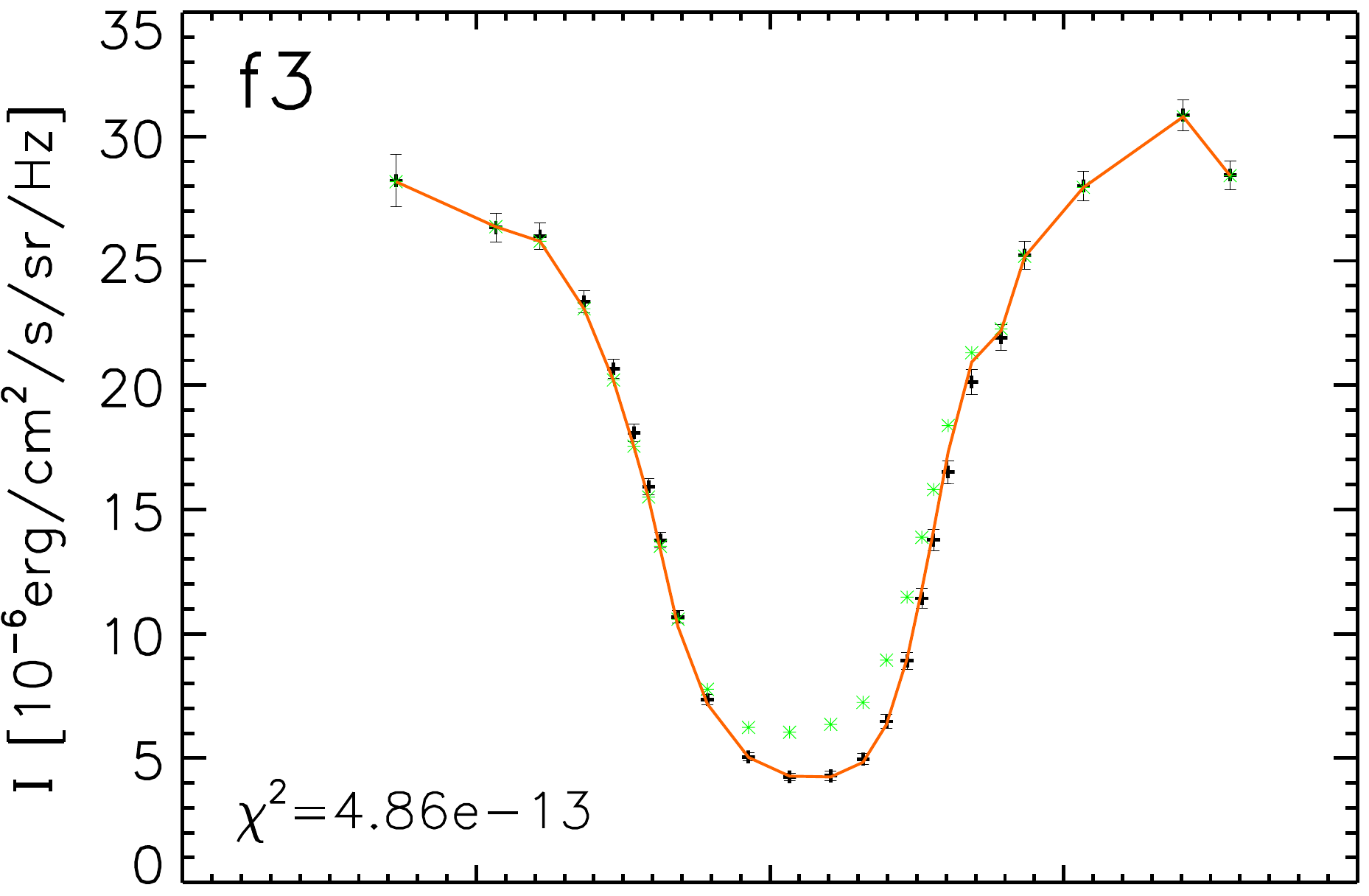}}}\ 
\parbox{0.11\columnwidth}{\phantom{1ex}}\
\parbox{0.89\columnwidth}{
\resizebox{0.89\columnwidth}{!}{
  \includegraphics{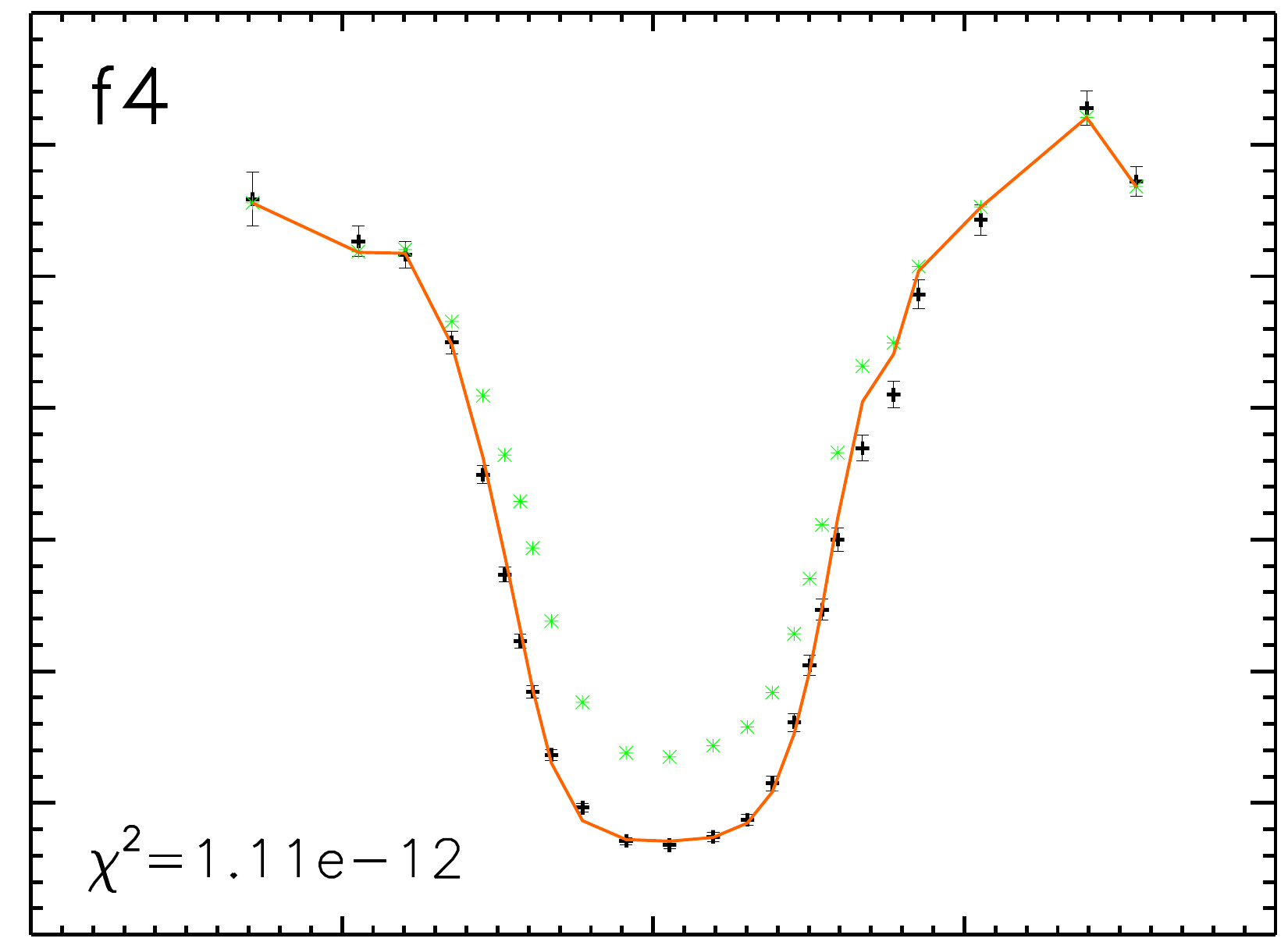}}
}\\
\parbox{\columnwidth}{
\resizebox{\columnwidth}{!}{
  \includegraphics{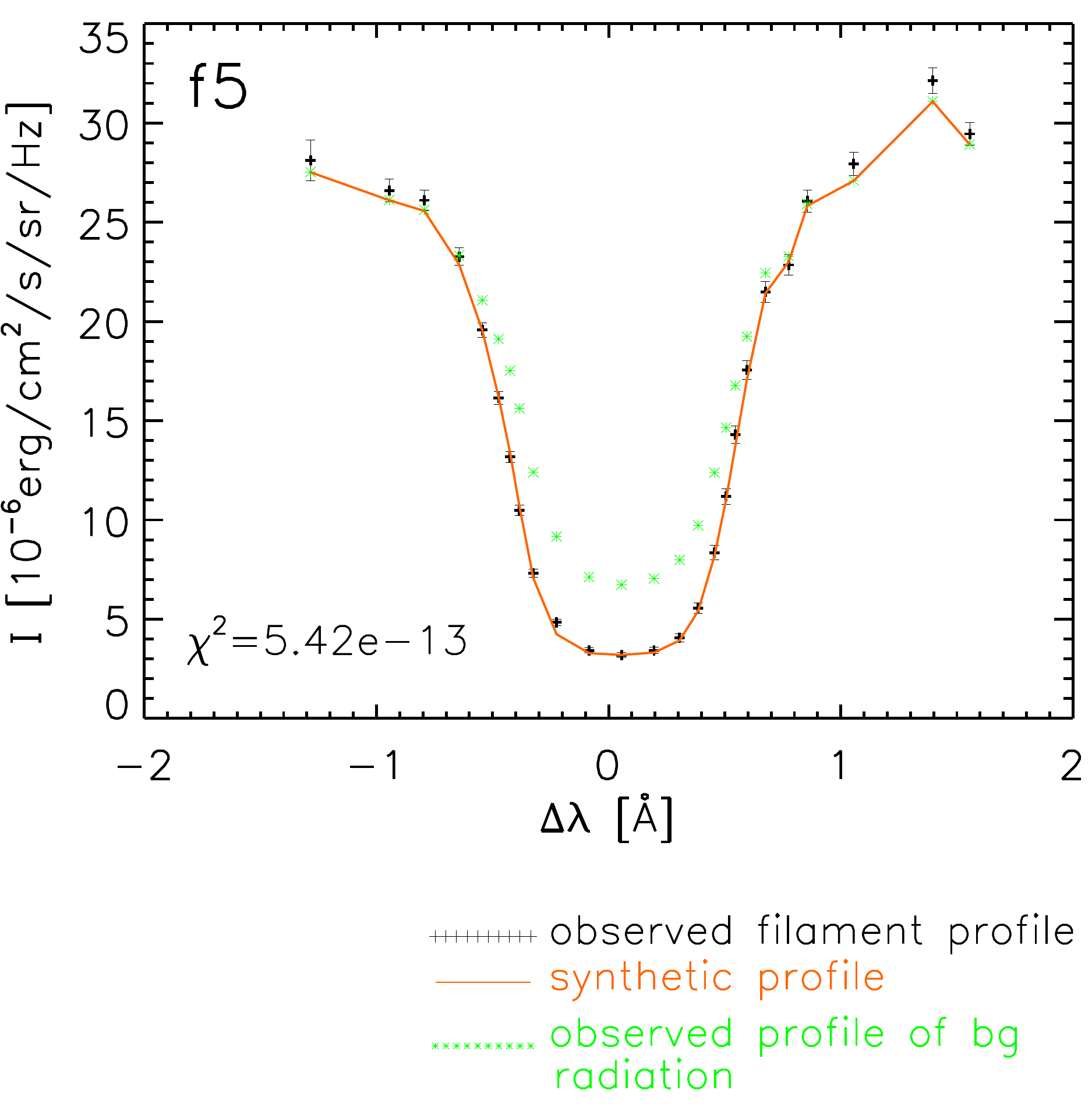}}}\  
\parbox{0.11\columnwidth}{\phantom{1ex}}\
\parbox{0.9\columnwidth}{
\resizebox{0.9\columnwidth}{0.995\columnwidth}{ 
  \includegraphics{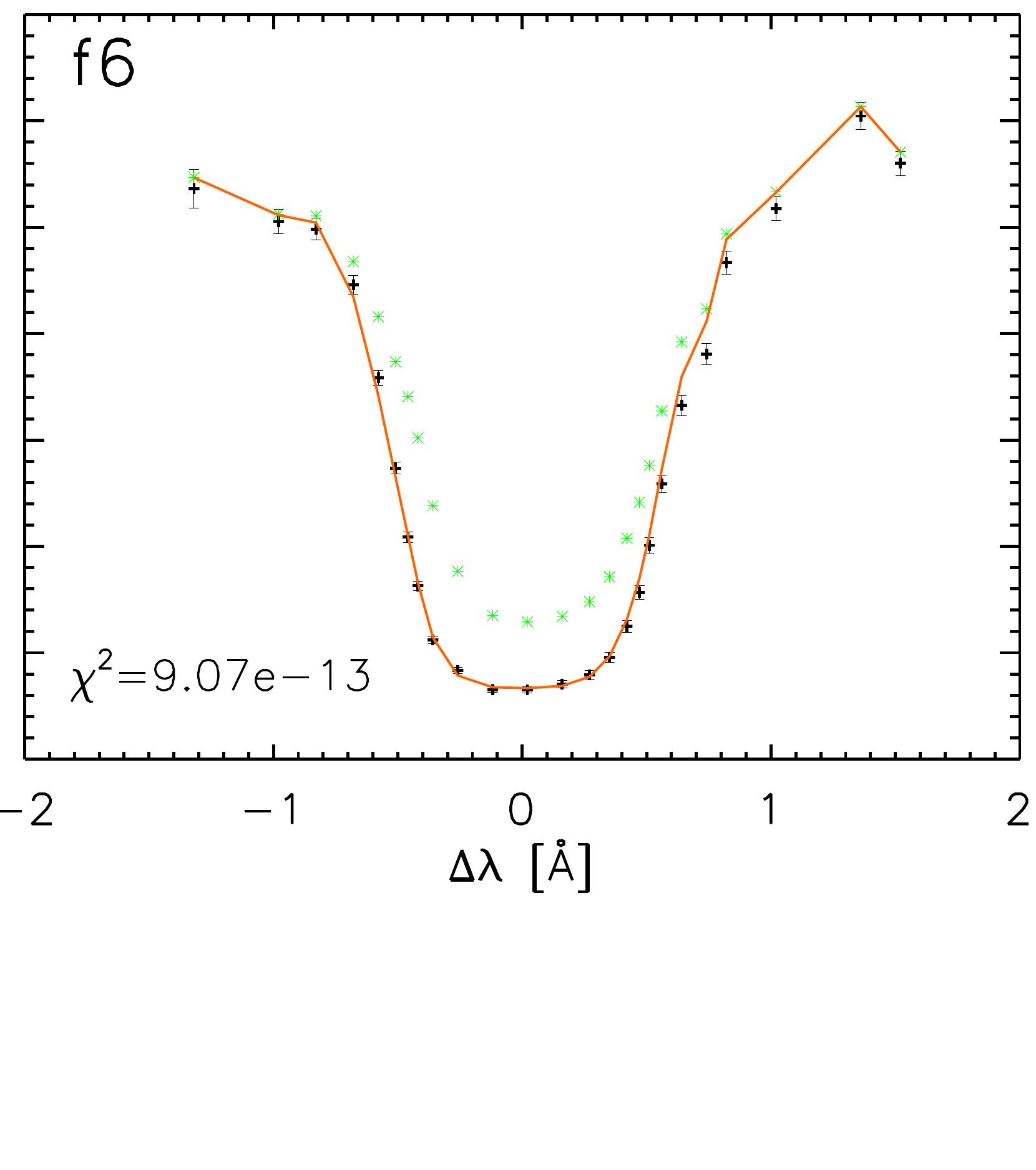}}} 
\caption{Comparison of the best synthetic profiles (red lines)
of the resulting models with observed profiles (black `+' symbols
with error bars) for the six areas f1\,--\,f6. The profiles from six
associated QS areas q1\,--\,q6 used as the background irradiation of the
slab in the formal solution are plotted by the green asteriscs.
In each plot the $\chi^2$ value of the synthetic profile fitting
to the observed one is also presented.}
\label{fig:profitting}
\end{figure*}

For the fitting of the observed profiles from each area
(f1\,--\,f6), a grid of 2D models was computed using the following
ranges of input parameters: temperature from $5,000$\,K to 
$19,000$\,K in increments of 2,000\,K; the gas pressure from
$0.05$ up to $0.80\,\mathrm{dyn}/\mathrm{cm}^2$ in increments of  
$0.02\,\mathrm{dyn}/\mathrm{cm}^2$; microturbulent velocity from
$0.5$ up to $14\,\mathrm{km/s}$ with the step of
$0.5\,\mathrm{km/s}$; the height of the filament
(Z\discretionary{-}{-}{-}dimension) from $10,000$ up to
$60,000\,\mathrm{km}$ in increments of $2,000\,\mathrm{km}$, and
the LOS component of the plasma\discretionary{-}{-}{-}flow
velocity $v_{\mathrm{LOS}}$ from $-7.5\,\mathrm{km/s}$ to 
$+1.5\,\mathrm{km/s}$ in increments of $0.2\,\mathrm{km/s}$.
Positive values of $v_{\mathrm{LOS}}$ mean direction towards the observer.

Approximation of the filament by the isothermal and isobaric slab is suitable
for the analysis of the H$\alpha$ spectral line because H$\alpha$
is predominantly formed within the cool and dense cores of
prominences and filaments. Therefore the influence of the hotter
prominence\discretionary{-}{-}{-}corona transition region (PCTR)
surrounding the cool cores is not significant and only
the cool cores need to be modelled. As the modelled 2D slabs
(see Fig.~\ref{fig:model_geom_scheme}) are narrow and observed
from the top, they thus represent the smaller\discretionary{-}{-}{-}scale.
Therefore we assumed the width of the modelled 2D slab 
(its X\discretionary{-}{-}{-}dimension) to be $1,000$\,km as was also adopted  
in modelling of the prominence fine\discretionary{-}{-}{-}structure threads, 
for example, by \citet{2005A&A...442..331H}.
%
%
%
%
\begin{table*}
\centering{
\caption{Additional properties of the best\discretionary{-}{-}{-}fit models with their uncertainties:
width of the projected slab, relative position -- the best\discretionary{-}{-}{-}fit LOS position 
relative to whole width of the projected slab (equal to zero at its south end and to 
$1$ at its north end), particle density, plasma density, ionisation degree of hydrogen, and the optical thickness $\tau_o^{(\mathrm{pos})}(H\alpha)$ in
the H$\alpha$ centre at the best\discretionary{-}{-}{-}fit LOS position.}
\label{table:resmodphysprop}
\begin{tabular}{ c | c c r r r r }
\hline \\[-2.3ex]
area & \multicolumn{1}{c}{projected width} & rel. position & \multicolumn{1}{c}{$n_{\mathrm{part}}$}   & \multicolumn{1}{c}{$\rho$}        &   \multicolumn{1}{c}{$i(\mathrm{H})$}    &
                                                                                                        \multicolumn{1}{c}{$\tau_o^{(\mathrm{pos})}(\mathrm{H\alpha})$}  \\
  & \multicolumn{1}{c}{[km]}           &    &    \multicolumn{1}{c}{$\left[\mathrm{cm}^{-3}\right]$} & \multicolumn{1}{c}{$\left[\mathrm{g/cm}^3\right]$} &                 &                       \\[0.5ex]
\hline \\[-1.7ex]
f1   &  $1,500$\,--\,$1,600$  &  $0.02$\,--\,$0.12$ & $2.10\times10^{11}\pm45\,\%$ & $3.26\times10^{-13}\pm51\,\%$ & $0.57\pm30\,\%$ &  $2.54\pm16\,\%$    \\
f2   &  $1,700$\,--\,$1,750$  &  $0.01$\,--\,$0.07$ & $1.52\times10^{11}\pm50\,\%$ & $2.45\times10^{-13}\pm49\,\%$ & $0.41\pm5\,\%\hspace{1.1ex}$  &  $2.02\pm5\,\%\hspace{1.1ex}$ \\
f3   &  $1,400$\,--\,$2,550$  &  $0.01$\,--\,$0.07$ & $4.32\times10^{11}\pm65\,\%$ & $8.46\times10^{-13}\pm68\,\%$ & $0.15\pm47\,\%$ &  $1.05\pm35\,\%$   \\  
f4   &  $1,500$\,--\,$1,650$  &  $0.03$\,--\,$0.25$ & $1.00\times10^{11}\pm51\,\%$ & $1.36\times10^{-13}\pm51\,\%$ & $0.80\pm11\,\%$ &  $3.88\pm17\,\%$    \\
f5   &  $1,750$\,--\,$2,200$  &  $0.03$\,--\,$0.15$ & $1.00\times10^{11}\pm51\,\%$ & $1.36\times10^{-13}\pm50\,\%$ & $0.75\pm3\,\%\hspace{1.1ex}$  &  $3.01\pm12\,\%$    \\
f6   &  $2,150$\,--\,$1,240$  &  $0.03$\,--\,$0.27$ & $8.40\times10^{10}\pm51\,\%$ & 
$1.07\times10^{-13}\pm51\,\%$ & $0.91\pm3\,\%\hspace{1.1ex}$ &  $3.89\pm13\,\%$
\\ \hline
\end{tabular}}
\end{table*}
%

Using the described filament model we obtained synthetic
H$\alpha$ profiles along the extent of the projection of the
2D slab onto the solar disk. The slab was inclined 
from the vertical LOS by the angle $\theta$. Hereafter, we
refer to the extent of this projection as the width of the
projected slab. We compared the synthetic H$\alpha$ profiles
obtained at all positions along the width of the projected slab
with the observed H$\alpha$ profiles. The 
best\discretionary{-}{-}{-}fit profile was determined by the 
$\chi^2$ minimum method, where $\chi^2$ was calculated as
\begin{equation}
\chi^2=\sum_i\,\frac{\left(I_{\mathrm{obs}}(\lambda_i)-I_{\mathrm{synth}}(\lambda_i)\right)^2}{\left(\rule[-1pt]{0ex}{2ex}\sigma_{\mathrm{I}}(\lambda_i)\right)^{2}}\,.
\label{eq:chisq}
\end{equation}
Here, $I_{\mathrm{obs}}(\lambda_i)$ and
$I_{\mathrm{synth}}(\lambda_i)$ are intensities of the observed
and the synthetic profiles, respectively. 
The quantity $\sigma_{\mathrm{I}}(\lambda_i)$ is the 
error of the observed intensity at the wavelength $\lambda_i$ calculated 
as multiplication of the relative error $\sigma_{\mathrm{rel}}(\lambda_i)$ 
(see Eq.~(\ref{eq:calibinterreq})\,) and the observed intensity. 
The position of LOS along the width of the projected slab from which 
the best\discretionary{-}{-}{-}fit synthetic profile was obtained, 
measured from the slab south end, was hereafter referred to as 
best\discretionary{-}{-}{-}fit LOS position. In this way we selected from
the grid of models the one that produced synthetic profiles
which had the best agreement with the observed profiles.
The comparison between the observed profiles from areas (f1 – f6) and their corresponding
$\chi^2$\discretionary{-}{-}{-}minimised synthetic profiles from the modelling is shown in
Fig.~\ref{fig:profitting}. 

In Table~\ref{table:resmodparams} we list the input parameters of the resulting models together
with their uncertainties. The uncertainties of the parameters were estimated by changing the
value of each parameter in both directions from the resulting model until the
best\discretionary{-}{-}{-}fit synthetic profile lies within the error bars of the observed profiles. 
In Table~\ref{table:resmodphysprop} we present an additional list of the physical parameters of the
resulting models. The 2nd and the 3rd columns are the width of the projected slab and the position
of the best\discretionary{-}{-}{-}fit LOS along the width of the projected slab, measured from its
south end (see an example of such a LOS in Fig.~\ref{fig:model_geom_scheme}, just the $\theta$ angle
in this figure is exaggerated for better intelligibility). 

The sensitivity of the derived best\discretionary{-}{-}{-}fit LOS position to the errors of the
measured spectral intensities is rather small. This was tested by fixing the model 
input variables and manually varying the best\discretionary{-}{-}{-}fit LOS position (such that the resulting 
profile remained within the error bars of the observed profiles shown in Fig.~\ref{fig:profitting}), we found that  
its uncertainties did not exceed $30$\,\% for all six filament areas. We then see that by varying the model input 
parameters, that were previously fixed, they have a more significant effect on the best\discretionary{-}{-}{-}fit LOS
position (up to $\pm80$\,\% of the value in the middle of each 'rel. position' range). Hence, the best\discretionary{-}{-}{-}fit
LOS position is more sensitive to the model inputs than to errors in the observed spectral intensities.  
In the next three columns, values of
the particle density (including free electrons, protons, neutral hydrogen, 
and neutral and ionised helium), plasma density, and the ionisation
degree of hydrogen are listed. We would like to point out that the listed values
of the densities and the ionisation degree are averaged over the
entire 2D slab and variations of these quantities within the slab
do not exceed $10\,\%$. The last column is the optical thickness
$\tau_o^{(\mathrm{pos})}(\mathrm{H\alpha})$ corresponding to the
position of the best\discretionary{-}{-}{-}fit LOS along the width
of the projected slab.
%
%
%
\section{Discussion}\label{s:discussion}
From the results listed in Tables~\ref{table:resmodparams} and
\ref{table:resmodphysprop} it is clear that the observed fragment of the filament
can be in general divided into two parts: region A at its SE side (areas
f1\,--\,f3 from Fig.~\ref{fig:sixareasimap}) and region B 
at its NW side (areas
f4\,--\,f6 from Fig.~\ref{fig:sixareasimap}). The observed H$\alpha$
profiles from region A (panels f1\,--\,f3 in Fig.~\ref{fig:profitting})
are less deep, not too broad and significantly asymmetric. Such profiles are
fitted by models with temperatures from $6,000$\,K to $10,000$\,K,
very high gas pressure of $0.20$\,--\,$0.40\,\mathrm{dyn/cm}^2$
and large plasma densities of the hydrogen and helium plasma
($2.5$\,--\,$8.5\times10^{-13}\,\mathrm{g/cm}^3$). The asymmetry
of these profiles results in well\discretionary{-}{-}{-}defined 
flows directed away from an observer with derived
LOS velocity components of $6$ to $7$\,km/s. In areas f1 and f3
(but not in f2) from region A we derived
rather fast unresolved plasma motions represented by the
micro\discretionary{-}{-}{-}turbulent velocities of $8$\,--\,$9$\,km/s.
On the other hand, the H$\alpha$ profiles detected in region B (areas 
f4\,--\,f6 in Fig.~\ref{fig:profitting}) are significantly deeper
and broader but more symmetric. Models that fit these profiles
indicate plasma with significantly higher temperatures
($11,000$\,--\,$14,000$\,K), lower gas pressure ($0.15\,\mathrm{dyn/cm}^2$)
and plasma densities of $1.1$\,--\,$1.4\times10^{-13}\,\mathrm{g/cm}^3$.
This part of the observed filament fragment exhibits smaller LOS velocities from  
$0$\,km/s to $-2.4$\,km/s. The derived micro\discretionary{-}{-}{-}turbulent velocities
are lower, $4$\,--\,$6$\,km/s.
Particle densities around $1\times10^{11}\,\mathrm{cm}^{-3}$ were obtained in all 
areas except the area f3 where a larger value of $4\times10^{11}\,\mathrm{cm}^{-3}$ was obtained. 
Values of $\sim\hspace{-3pt}10^{11}\,\mathrm{cm}^{-3}$ are comparable to those obtained by \citet{2006A&A...459..651S}
for another quiescent filament observed in the hydrogen Lyman series.
Comparing the particle densities to plasma densities concerning the six filament areas indicates   
that for the areas f1 and f4\,--\,f6 there is a linear relation between the two quantities.
Only for areas f2 and f3, deviations from this linear relation to larger plasma densities occur.
These deviations are caused by a lower degree of ionisation of hydrogen in these two areas because 
there was smaller fraction of free electrons (which were included in particle density but their
contribution to plasma density is negligible due to their very low mass) among other particles
(protons, neutral hydrogen, and atoms and ions of helium) than in other filament areas. The deviation
to larger plasma densities is remarkable mainly in the area f3 where a much lower ionisation degree of hydrogen
than in other five filament areas was estimated. The observed filament fragment is in
general geometrically rather extended along the Z\discretionary{-}{-}{-}axis 
with the derived vertical size of $22,000$\,--\,$38,000$\,km. 
For the filament observed by the AIA instrument few days earlier as a prominence at the limb a height of 
approximately $20,000$\,km was measured what is in quite good agreement with values of the filament vertical size 
obtained from the modelling. 
In summary, the derived parameters indicate that region A was cooler,
more dense, and more dynamic while in region B it was hotter, less dense, and more quiescent.
%
\newlength{\figsevenl}
\setlength{\figsevenl}{0.25\textwidth}
\newlength{\figsevenlaa}
\setlength{\figsevenlaa}{\textwidth-\figsevenl}
\newcommand{\factormidpanel}{0.50}
\newlength{\figsevenm}
\setlength{\figsevenm}{\factormidpanel\figsevenlaa}
\newlength{\figsevenr}
\setlength{\figsevenr}{\figsevenlaa-\figsevenm}
\begin{figure*}
\parbox[][][t]{\figsevenl}{
\resizebox{\figsevenl}{!}{
  \includegraphics{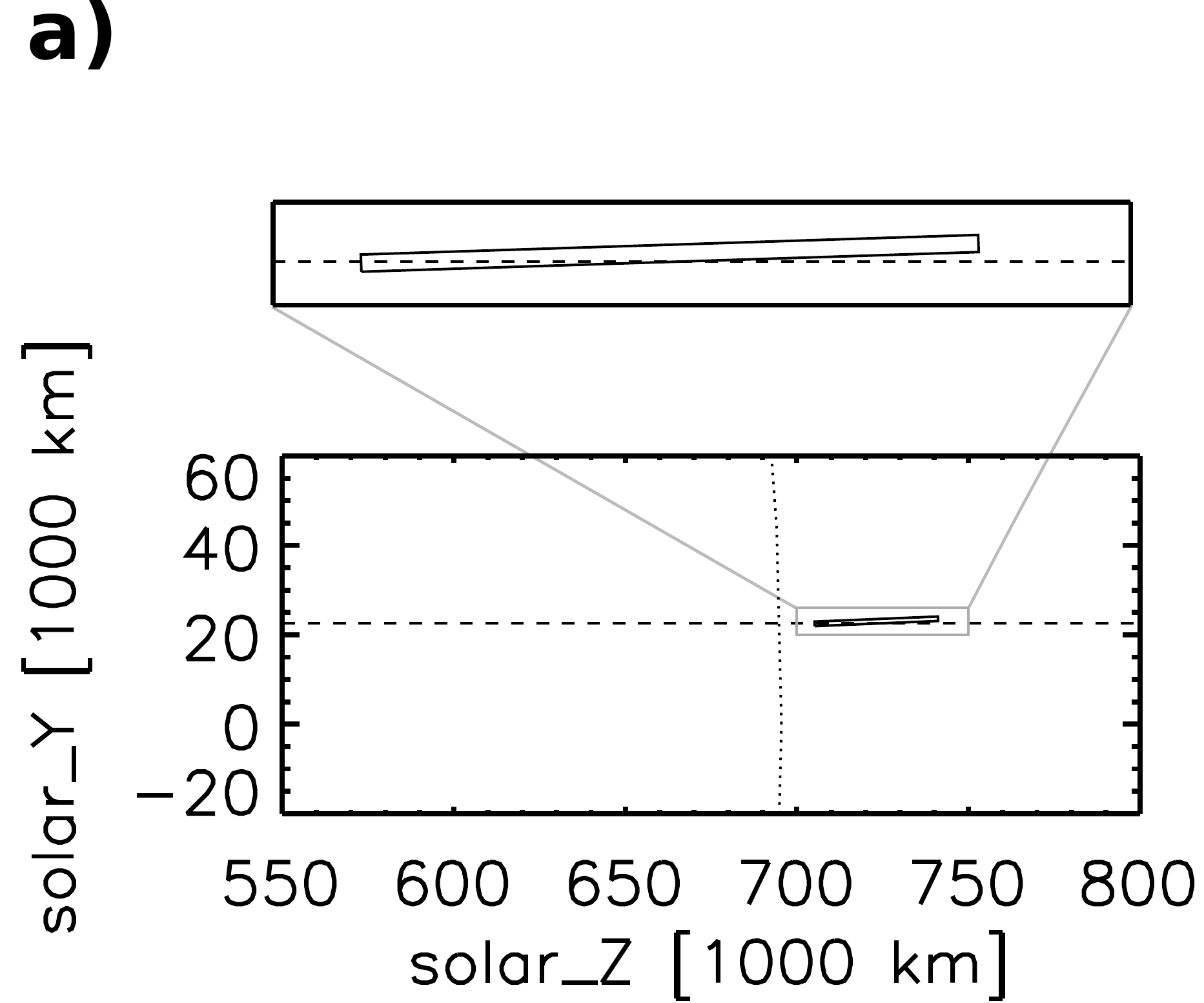}}
\phantom{1ex}\\[-0.06\figsevenl]}\
\parbox[][][t]{\figsevenm}{
\resizebox{\figsevenm}{!}{
  \includegraphics{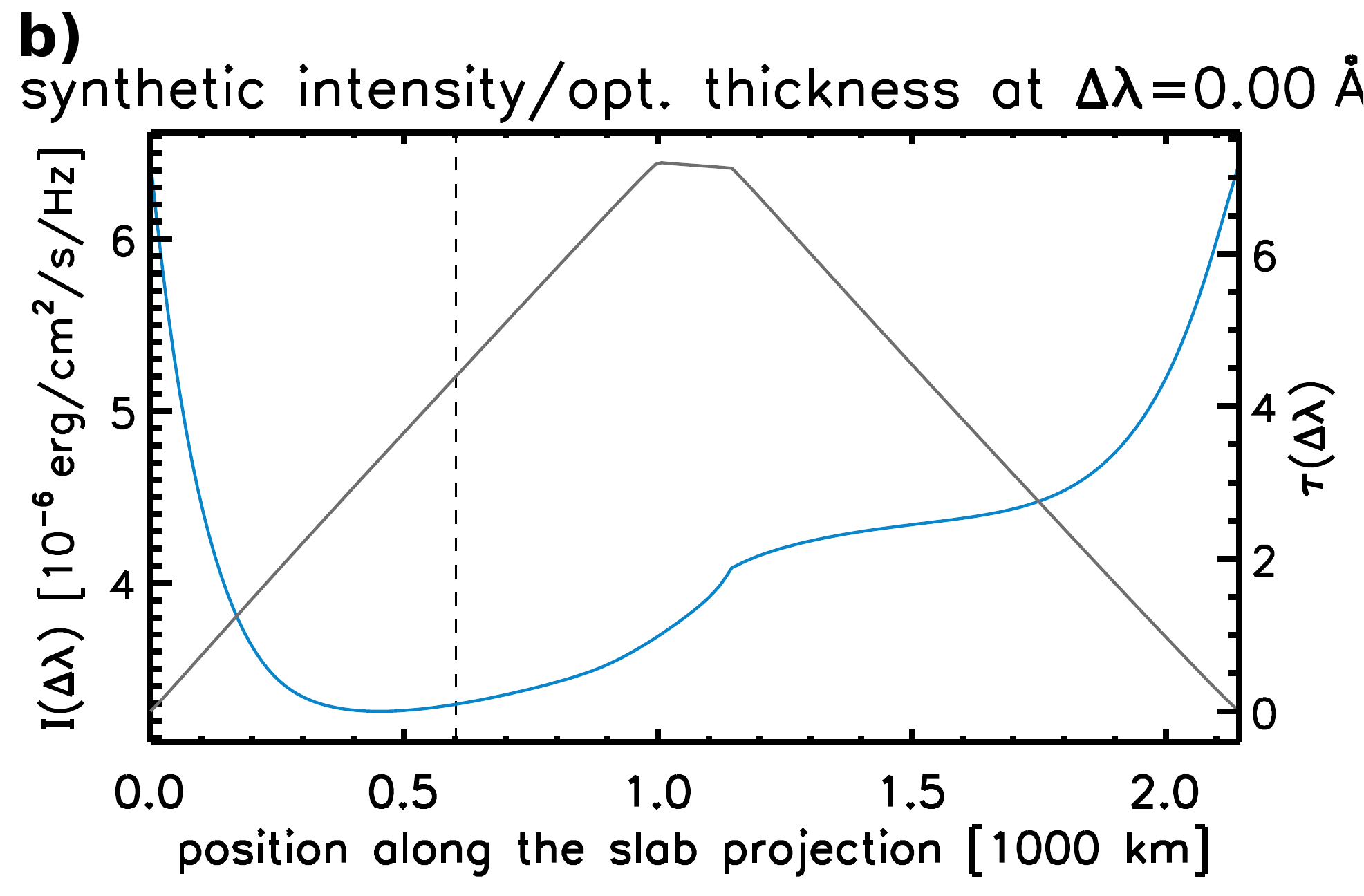}}}\ 
\parbox[][][t]{\figsevenr}{
\resizebox{\figsevenr}{!}{
  \includegraphics{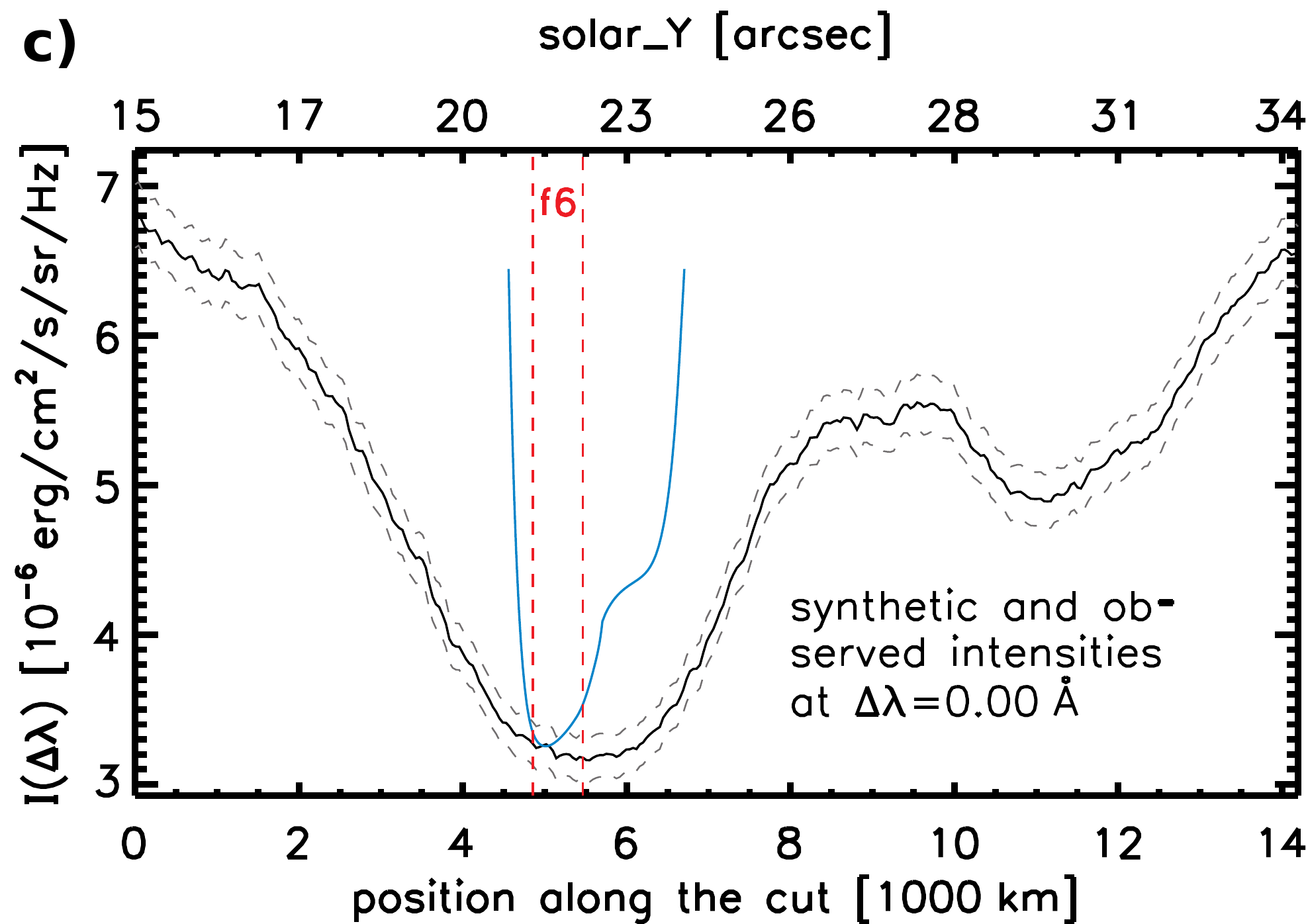}}}\\[4ex] 
\parbox[][][t]{\figsevenl}{
\resizebox{\figsevenl}{!}{
  \includegraphics{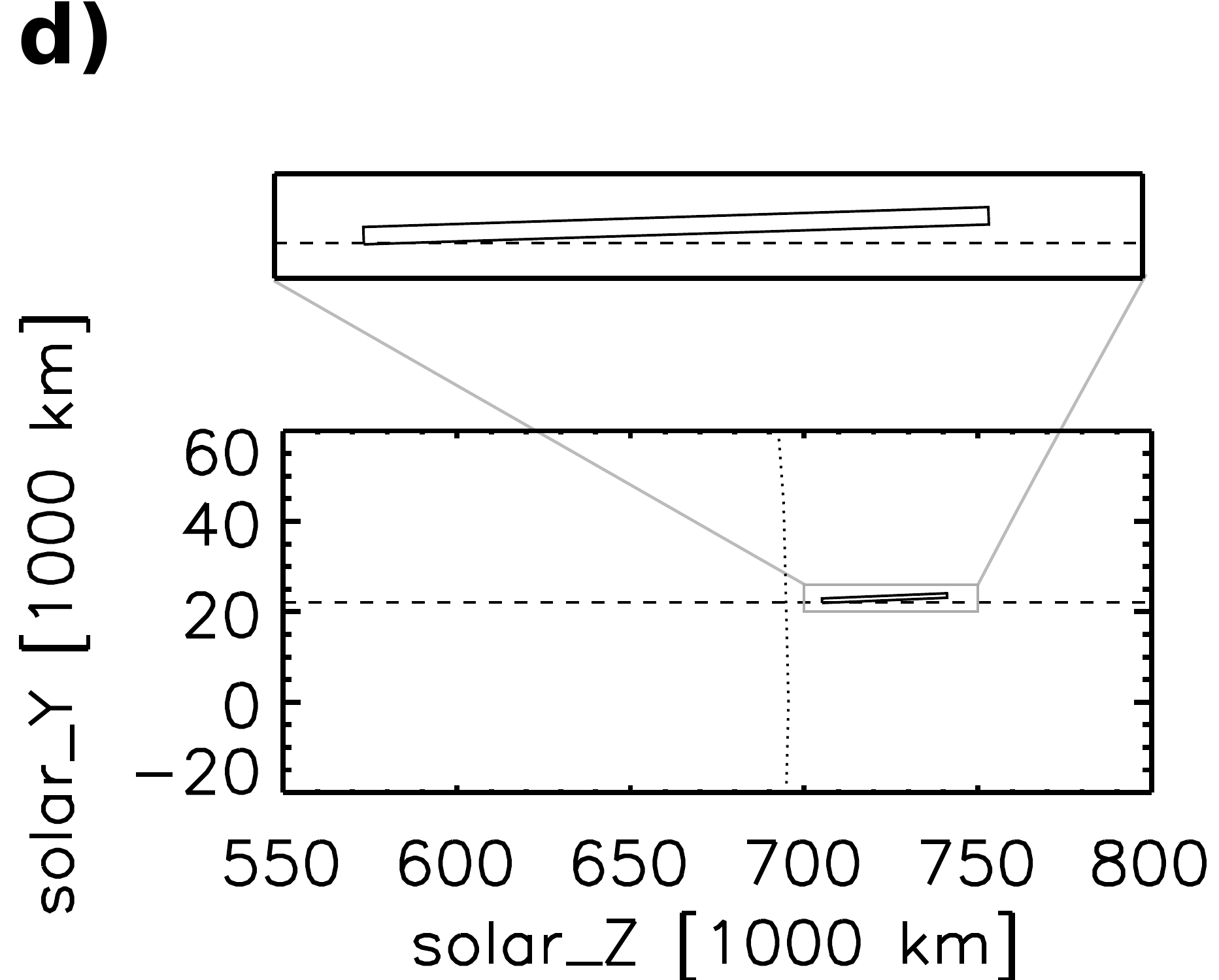}}   
\phantom{1ex}\\[0.197\figsevenl]}\
\parbox[][][t]{\figsevenm}{
\resizebox{\figsevenm}{!}{
  \includegraphics{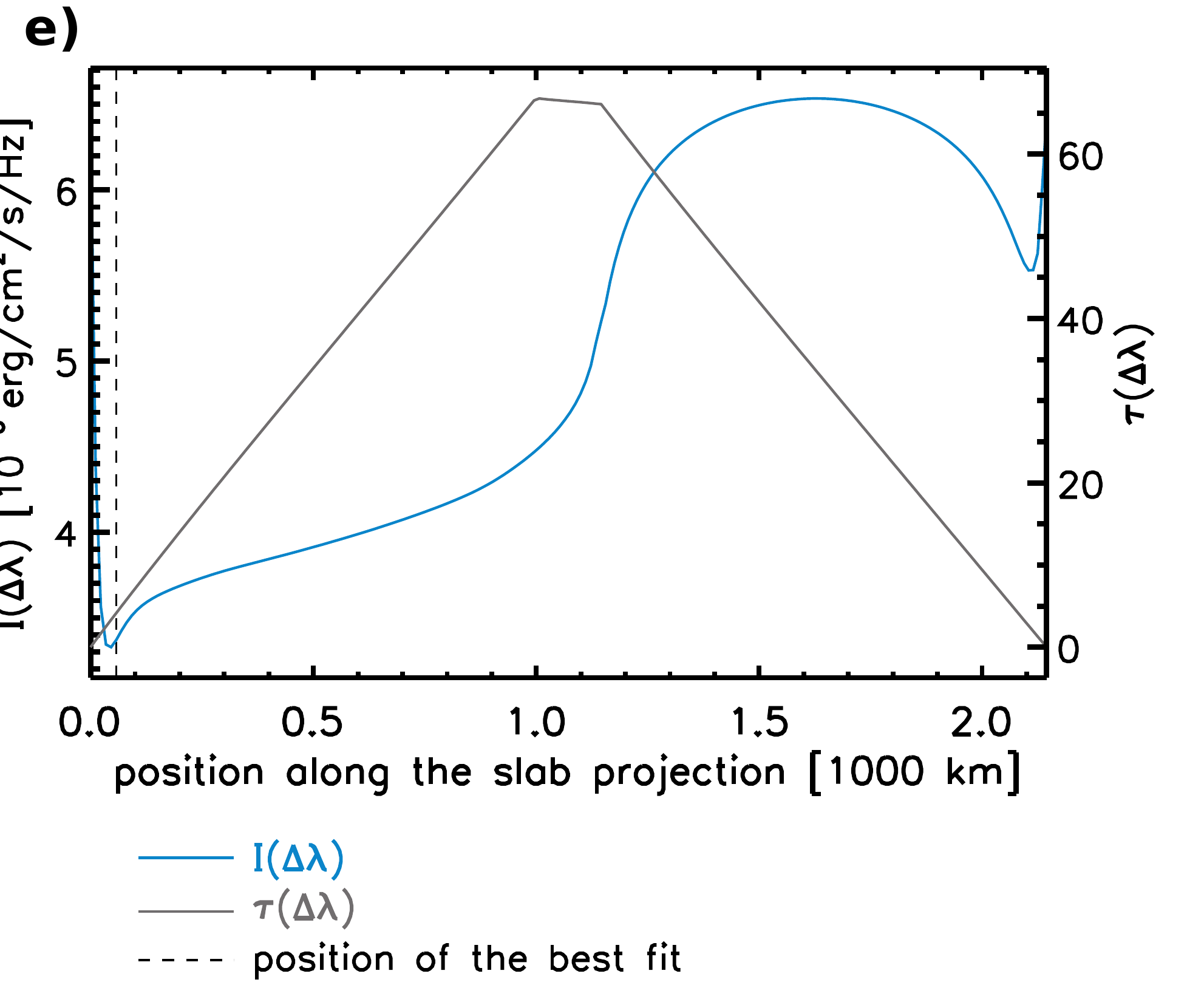}}}\ 
\parbox[][][t]{\figsevenr}{
\resizebox{\figsevenr}{0.940\figsevenr}{
  \includegraphics{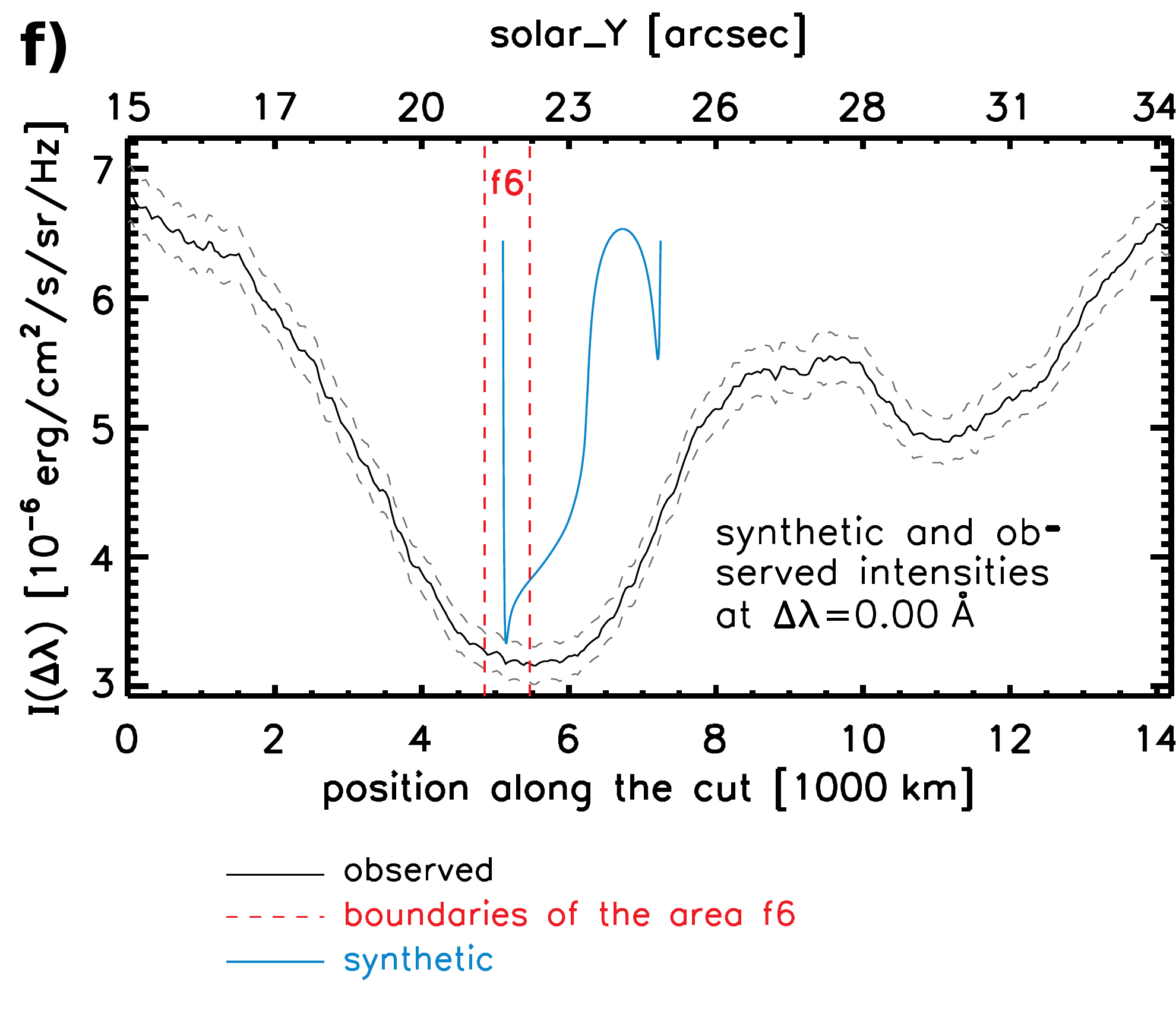}}}
\caption{Example of behaviour of the model and comparison of
resulted synthetic intensities with observations for area f6. In
the upper row we show the results of a model with the gas pressure
of $0.08\,\mathrm{dyn/cm}^2$ and in the bottom row the results of
a model with the gas pressure of $0.25\,\mathrm{dyn/cm}^2$. The
spectral intensities $I(\lambda)$ in the plots in panels
\texttt{b}, \texttt{c}, \texttt{e}, and \texttt{f} are expressed in
units of $10^{-6}\,\mathrm{erg/cm}^2\mathrm{/s/Hz}$.
Comparing panels \texttt{a}, \texttt{b}, and \texttt{c}
with panels \texttt{d}, \texttt{e}, and \texttt{f} one can see that
for results of the modelling obtained with rather different values of gas pressure,
there exist synthetic profiles for both models which agree well with the observed
profile within errors of its measured spectral intensities (in panels \texttt{c} and \texttt{f}
shown only for $\Delta\lambda$=$0$\,\AA), although those synthetic profiles occur
in radiation emergent from different positions along the projected slab.
More details about the plots shown in the figure are given in the text of
section \ref{s:uncertain}.} 
\label{fig:discussion1}
\end{figure*}
%
\subsection{Discussion of background radiation}
To support the reliability of the modelling, the influence of the
selection of the background\discretionary{-}{-}{-}radiation
profiles used in the formal radiative transfer solution on the
results should be discussed. 
As it was not possible to directly obtain the background radiation
profiles at the surface under the filament, 
we approximated such a background by selecting 
areas outside the filament (q1\,--\,q6), which correspond to the
areas f1\,--\,f6 in the filament (see Fig.~\ref{fig:sixareasimap}).
The areas q1\,--\,q6 were selected as close as possible to the
corresponding filament areas to obtain a background radiation
profile that was as realistic as possible. A 
good way to test the influence of the selected
background\discretionary{-}{-}{-}radiation areas on the modelling
is to compare the results using the areas q1\,--\,q6 with results
using the average QS profile used for the data
calibration (see Sect.~\ref{s:datacalib}). For the QS profile an
extended area with dimensions of
$10\,\mathrm{arcsec}\,\times\,40\,\mathrm{arcsec}$ located
relatively far from the filament (approx. $20$\,arcsec, see
Fig.~\ref{fig:context}\texttt{c}) was selected to minimise any influence of
plasma flows occurring in the vicinity of the filament. This way,
any asymmetry or Doppler shift was eliminated from the QS
profile (see panel \texttt{a}\,--\,\texttt{d} in Fig.~\ref{fig:pre-method}).

To understand the asymmetry of an absorption profile more clearly, 
we can define $r_{\mathrm{asym}}$ as a ratio between intensities
$I(\lambda_{\mathrm{min}}-0.5\,\AA)$ and
$I(\lambda_{\mathrm{min}}+0.5\,\AA)$ at wavelengths
$\pm0.5\,\mathrm{\AA}$ from the wavelength
$\lambda_{\mathrm{min}}$ where the intensity profile reaches the
minimum:
\begin{equation}
r_{\mathrm{asym}}=\frac{I(\lambda_{\mathrm{min}}-0.5\,\AA)}{I(\lambda_{\mathrm{min}}+0.5\,\AA)}\,.
\label{eq:prof_asymmetry}
\end{equation}
This ratio is equal to unity for the QS profile and close to unity
for profiles from areas q1\,--\,q5 ($0.96$\,--\,$1.07$).
Only the background radiation profile from area q6 is more
asymmetric with $r_{\mathrm{asym}}=1.23$. The difference of
the $\lambda_{\mathrm{min}}$ positions between profiles from areas
q1\,--\,q5, and the QS profile are negligible ($\ll 0.1\,\AA$).
While the q1, q2, q4, and q5 profiles have a similar depth to
the QS profile, the profiles from areas q3 and q6 are considerably
deeper, with minimum intensities 12\% and 6\% lower, respectively.

To assess the influence of the choice of the background radiation
on the modelling results, we used the model parameters derived as
the best fit for each area f1\,--\,f6 and substituted only the
background profiles. In areas f1, f2, and f4, only small differences
in the resulting parameters occur -- up to $18\,\%$ for the optical thickness
$\tau_o^{(\mathrm{pos})}(\mathrm{H\alpha})$ and for other
quantities only up to $7\,\%$. The situation in the area f5 is
different -- differences for some resulting quantities are much
larger, for instance, $62\,\%$ in $v_{\mathrm{LOS}}$ or $37\,\%$ in
$\tau_o^{(\mathrm{pos})}(\mathrm{H\alpha})$. This might be caused
by f5 being located between the two regions -- between the
dynamic and the heated parts of the observed filament fragment   
which could cause numerical instability to affect the results
(i.e. small changes in the modelling can cause larger changes in the results).
However, it is important that the change in the plasma density
is also small in area f5. Despite the fact that the q3 and q6
profiles are deeper then the QS profile, most of the properties
derived in the areas f3 and f6 when the QS profile is used change
by only a few \%. Larger differences occur only in
$\tau_o^{(\mathrm{pos})}(\mathrm{H\alpha})$ which increases by up
to 63 \%. The most significant difference occurs in the area f6,
where we obtained $v_{\mathrm{LOS}}$ of $-0.7\,\mathrm{km/s}$
instead of $-0.1\,\mathrm{km/s}$ (difference of 85\,\%). This is due to the significant
asymmetry of the q6 profile mentioned above.

This analysis shows that the selection of the background radiation
from smaller areas close to the studied areas within the filament does
not significantly influence the parameters derived by the modelling.
However, we would argue that in order to achieve a higher degree of accuracy,
it is better to use such local background radiation profiles
instead of an averaged QS profile. This may be important
especially for the derivation of the LOS velocities, as can be
seen in the case of the area f6.
%
\subsection{Assessment of uncertainties of derived parameters}\label{s:uncertain}
The temperature, gas pressure, and micro\discretionary{-}{-}{-}turbulence velocity are the main parameters 
responsible for the width and the depth of the resulting H$\alpha$ profiles. From these 
it is the temperature which is derived with the least uncertainty (see Table~\ref{table:resmodparams}).  
When studying the influence of the temperature on the shape of the synthetic profiles we found that for low
values of the gas pressure (lower than $\sim\hspace{-2pt}0.08\,\mathrm{dyn/cm}^2$)
and for a temperature of $7,000\,\mathrm{K}$ our 2D model produces synthetic
profiles which are deep enough but much narrower than those observed in the
studied filament. With an increasing temperature, the synthetic profiles become slightly wider
without changin their depth, although they do not widen enough to fit the observed profiles.
Finally, when temperature is increased above $10,000\,\mathrm{K}$, profiles become shallower and
are no longer deep enough to match the depth of the observed profiles. On the other
hand, for higher gas pressure the profile depth increases with a temperature of up to around 
$11,000\,\mathrm{K}$, the synthetic profiles are then wide enough to accurately represent 
the observed profiles from the region A. Further increasing the model temperature 
causes a considerable broadening of the profiles resulting in both broad and
deep profiles that fit those observed from the region B. It is not possible to fit these broad
and deep profiles even by increasing the micro\discretionary{-}{-}{-}turbulence velocity
instead of the temperature while fixing the high gas pressure. This is because
$v_{\mathrm{MT}}$ makes profiles shallower while it broadens them.
Therefore, higher temperatures are inevitable for the region B.
An increase of the temperature
above a critical value (e.g. $15\,000\,\mathrm{K}$ for pressure of
$0.15\,\mathrm{g/cm}^2$) also makes the profiles shallower. We can
thus conclude that for the values of the gas pressure obtained for
all six areas of the filament, the shape of all observed H$\alpha$
profiles is strongly sensitive to the temperature. Therefore, the
uncertainties in the derived temperature values are small (up to
$11\,\%$) and the main source of these uncertainties are the
errors in the measured spectral intensities.  
%
\begin{figure*}
\parbox{0.46\textwidth}{  
\resizebox{0.46\textwidth}{!}{
\includegraphics{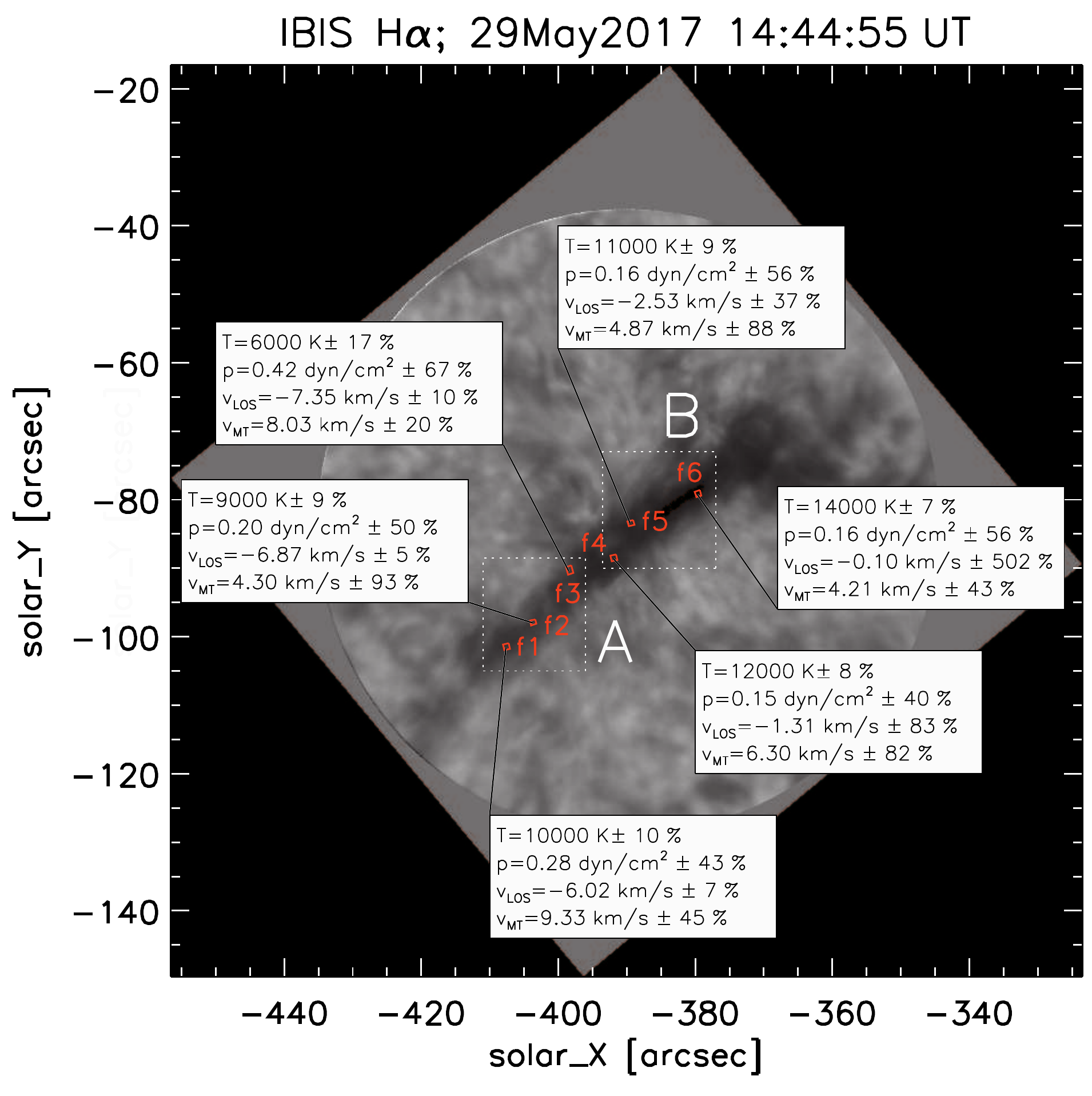}}}  
\parbox{0.46\textwidth}{  
\resizebox{0.46\textwidth}{!}{
\includegraphics{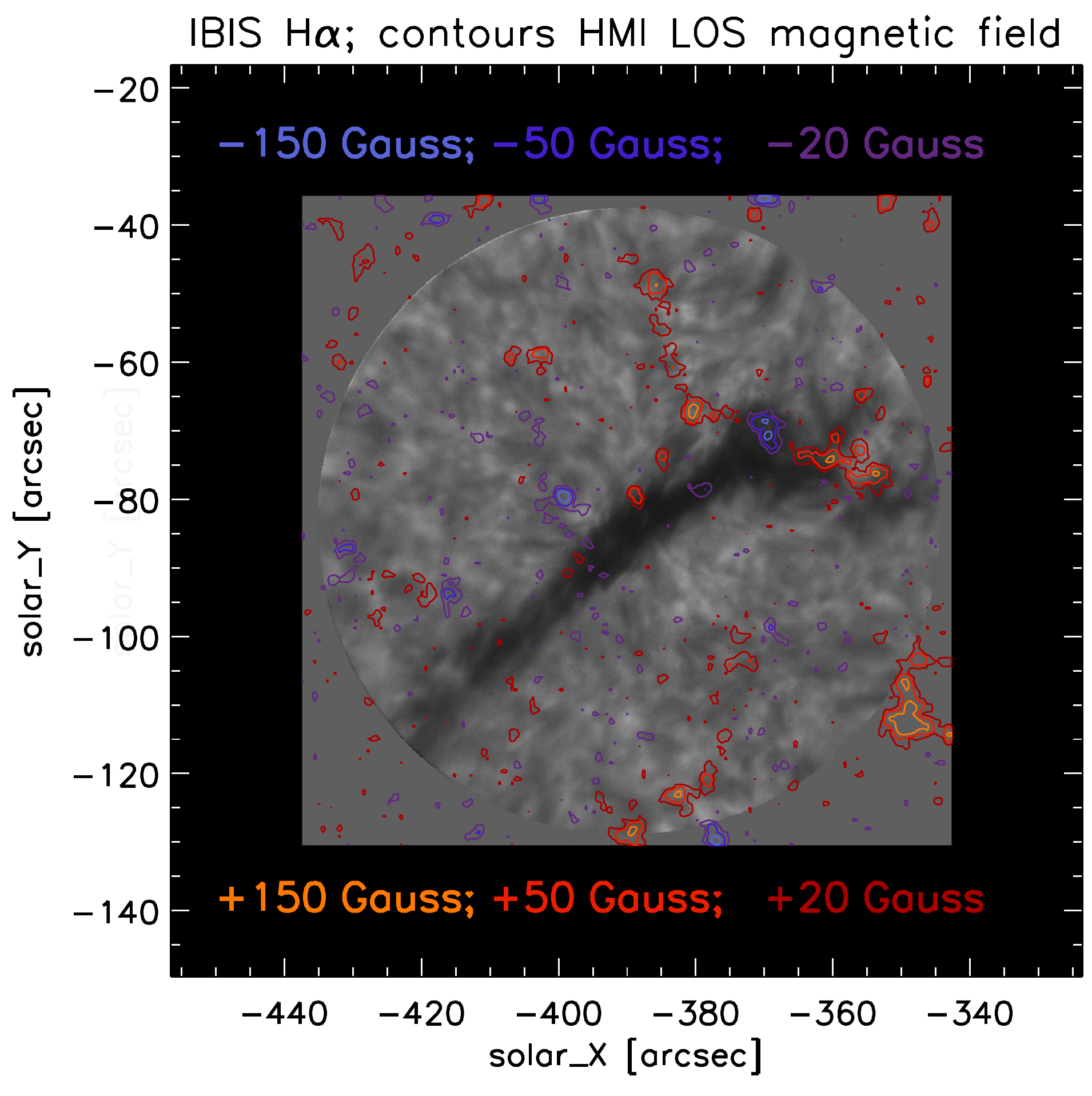}}} 
\caption{Intensity map of the observed fragment of the filament with schematically shown plasma
parameters derived for individual areas f1\,--\,f6 using the forward modelling shown in the left panel.
In the right panel, the intensity map overplotted with contours of the LOS magnetic field of
$-150$, $-50$, $-25$, $+25$, $+50$, and $+150$\,Gauss measured by the SDO/HMI instrument.}
\label{fig:intmapresults}
\end{figure*}

The uncertainties of the derived gas pressure and
micro\discretionary{-}{-}{-}turbulence velocity values, and
subsequently of the particle density and plasma density, are significantly
larger than in the case of the temperature. To explain the causes
of these uncertainties, we constructed two additional models
for the area f6. In these models, we altered the gas pressure
to be $0.08$ and $0.25\,\mathrm{dyn/cm}^2$, respectively. These
values remained within the uncertainty for the best\discretionary{-}{-}{-}fit
model for the area f6 (see Table~\ref{table:resmodparams}). In
Fig.~\ref{fig:discussion1}, we show the results of these two 
alternative models -- in the first row (panels \texttt{a} to
\texttt{c}) the model with the gas pressure of
$0.08\,\mathrm{dyn/cm}^2$ and in the second row (panels
\texttt{d} to \texttt{f}) the model with the gas pressure of
$0.25\,\mathrm{dyn/cm}^2$. The first panel in each row shows the
geometric scheme of the model with the LOS (dashed line) passing
through the slab embedded in the Z\discretionary{-}{-}{-}Y computational
domain as shown in Fig.~\ref{fig:model_geom_scheme}. The dotted
line denotes the solar surface. The second panel in each row of
Fig.~\ref{fig:discussion1} shows the variation of the intensity
(blue line) and the optical thickness (black line) at the
H$\alpha$ line centre. These variations are displayed along the
projection of the modelled 2D slab onto the solar surface. The
positions of the best fit are marked by dashed vertical lines. The
third panel in each row of Fig.~\ref{fig:discussion1} shows the
distribution of the observed intensity at the H$\alpha$ line
centre along a cut through the observed fragment of the filament 
from the solar S to the solar N. The cut passes through the centre of the area f6.
The intensity is plotted with the solid black line and its errors
with thin dashed black lines. The model with the lower pressure
(upper row of Fig.~\ref{fig:discussion1}) contains plasma with
lower density than the higher\discretionary{-}{-}{-}pressure model.
Therefore, to produce the same intensity the LOS in the case of the
lower\discretionary{-}{-}{-}pressure model has to cover a larger
distance within the modelled slab. Interestingly, even though the
optical thickness values along the slab projection (see panels
\texttt{b} and \texttt{e} of Fig.~\ref{fig:discussion1}) differ
significantly between the models, the $\tau$ values at the position
of the best fit are similar. This means that the column
mass along these two lines of sight is also similar.

Another interesting aspect is demonstrated by panels \texttt{c}
and \texttt{f} of Fig.~\ref{fig:discussion1}. In these plots we
show in the solid blue lines the distributions of the synthetic
H$\alpha$ line centre intensities (the same as in panels
\texttt{b} to \texttt{e}). For both models the synthetic intensity
reaches values between the errors of the observed intensities
within the borders of the area f6 marked by dashed red vertical
lines. Panels \texttt{c} and \texttt{f} also show that the width
of the projection of the modelled slab represented by the extent
of the blue profiles is clearly narrower than the dip in the observed
intensities which corresponds to the width of the observed
fragment of the filament. This means that we would need
to place multiple 2D slabs next to each other to simulate the entire
extent of the filament width. We will develop such a
multi\discretionary{-}{-}{-}slab model in the future.

Another source of significant uncertainties which are
included in the uncertainty budgets listed in Tables
\ref{table:resmodparams} and \ref{table:resmodphysprop}, is the
fact that we cannot exactly determine the position along the width
of the projected slab at which the best fit occurs. The
determination of this position affects the length of the
cross-section of a LOS and the modelled inclined 2D slab. This
subsequently affects the derived values of the pressure, density, 
and ionisation degree, but also the micro\discretionary{-}{-}{-}turbulent velocity.
However, the column mass derived at the different positions along
the width of the projected slab does not vary significantly.
We want to emphasise here that the best fit between the synthetic
and observed H$\alpha$ profiles was obtained at all wavelengths,
not only in the line centre.

An additional source of uncertainty, beyond that 
explicitly addressed here, lies in 
fact that we have only the H$\alpha$ line
available for this analysis. In the H$\alpha$
line the filament is much less optically thick than for example in
the Ly$\alpha$ or Ly$\beta$ lines. The optical thickness inside a
filament in the Ly$\alpha$ line centre was shown to be of the
order of $10^6$ by \citet{2006A&A...459..651S}. On the other hand,
$\tau_o^{(\mathrm{pos})}(\mathrm{H\alpha})$ in the H$\alpha$ line
centre in the filament studied here was derived to be up to $4$.
This means that in the case of the H$\alpha$ line, most of the
plasma along the LOS contributes to the resulting intensity. In
the Ly$\alpha$ case, only a very limited part of the plasma along
the LOS (the plasma nearest to the observer) contributes to the
resulting intensity. This can be seen for example in plots  of the
contribution function in Fig.~9 of \citet{2012SoPh..281..707S}.
Indeed, it is easier to determine the plasma conditions within a
localised area than in the plasma extended along the LOS. 
Unfortunately we are not able to quantify the   
influence of this effect on uncertainties of the results of the modelling, 
nevertheless we want to note that these results are influenced 
also by this effect.
%
\section{Conclusions}\label{s:Conclusions}
In the present paper we analysed the H$\alpha$ spectral
observations of a fragment of the large filament obtained by the DST/IBIS
on May 29, 2017. For the analysis we used a 2D non\discretionary{-}{-}{-}LTE filament
model which enabled us to produce synthetic H$\alpha$ spectra
that were compared with the observations. Using this forward
modelling technique, we derived the thermodynamic properties of
plasma in selected positions within the observed filament fragment 
(see left panel of Fig.~\ref{fig:intmapresults}). Our analysis shows that the observed filament fragment 
can be broadly divided into two parts. One part marked as region A (areas f1\,--\,f3)
is cooler, denser, and more dynamic. The derived temperatures range from $6,000\,\mathrm{K}$ to $10,000\,\mathrm{K}$
and the gas pressure is from $0.2\,\mathrm{dyn/cm}^{2}$ to $0.4\,\mathrm{dyn/cm}^{2}$. The more dynamic nature
of this region is characterised by the LOS velocities from $-6\,\mathrm{km/s}$ to $-7\,\mathrm{km/s}$ (negative velocities  
mean receding flows) and the
micro\discretionary{-}{-}{-}turbulent velocities of $8$\,--\,$9\,\mathrm{km/s}$. On the other hand, the 
other part marked as region B is hotter, less dense, and more quiescent. The
derived temperatures are $11,000$\,--\,$14,000\,\mathrm{K}$ and the gas
pressure is around $0.15\,\mathrm{dyn/cm^{2}}$. The derived LOS velocities
are from $-2.4\,\mathrm{km/s}$ to $0\,\mathrm{km/s}$ and the micro\discretionary{-}{-}{-}turbulent velocities are
$4$\,--\,$6\,\mathrm{km/s}$. We want to emphasise here that the broad and deep
H$\alpha$ profiles observed in this region cannot be reproduced by significantly lower temperatures using our 2D model.
We would like to point out that higher temperatures are also found by \citet{2019A&A...624A..72Z} along with some 
mass\discretionary{-}{-}{-}draining for a different prominence observed on 28 May 2014 on the NW limb during its  
activation phase a couple of hours before an eruption.  

The distinct hooked shape of the filament suggests that a barb or one of its footpoints was rooted within
the FOV. To explore this, we can first assume that the motions of plasma within the filament were largely 
tied to the topology of the host magnetic field. Then, as the filament did not display any bulk oscillatory motions,
we can assume that the observed LOS velocities correspond to motions of plasma that were aligned with the host magnetic
field. Region A contained larger LOS velocities when compared with region B, this can be interpreted as the
magnetic field associated with region A being more aligned with the LOS than region B.
The velocities were directed away from an observer, this can be further interpreted as mass flowing into the body
of the filament. For such a supposed topology of the magnetic field, this suggests a curvature in
the filament towards the surface moving from region A towards region B. This is in agreement with the conclusions of
\citep{Jenkins:2019b} in which the authors use the combination of a (double) Beckers Cloud Model
\citep[e.g.][]{tziotziou:2007}, a LOS\discretionary{-}{-}{-}projected HAZEL inversion from 
\citet{Wang:2019}, and both local and non\discretionary{-}{-}{-}local thermodynamic equilibrium inversions
\citep[e.g.][]{Beck:2015, Beck:2019} to study plasma flows and reconstruct the 3D topology of the filament
footpoint. 

Specifically, such a conclusion -- region A contained a magnetic field more parallel to the LOS than region B
that contained a magnetic field more perpendicular -- is supported by the LOS\discretionary{-}{-}{-}projected
HAZEL inversions of spectropolarimetric observations taken in the \ion{He}{i} infrared triplet
\citep{2008ApJ...683..542A}. And in the hook\discretionary{-}{-}{-}shaped structure located NW from region B the field 
was more vertical again, also suggested by the coincidence  
of concentrations of the LOS surface magnetic flux at approximately ($-380$\,arcsec in
solar~X, $-75$\,arcsec in solar~Y), observed by the Helioseismic and Magnetic Imager \citep[HMI; ][]{2012SoPh..275..207S}
instrument on board the SDO satellite and shown in the right panel of Fig.~\ref{fig:intmapresults}. 
The results of the modelling presented here are affected by several sources of uncertainties and we 
addressed and quantified the main contributors in Sect.~\ref{s:uncertain}. While the uncertainties in the
determination of some of the parameters may be significant, these do not affect the broad conclusions stated
above. To reduce the uncertainties affecting our forward modelling we plan to develop the method further to
use a more sophisticated 2D model with multiple fine structures along a LOS.
%
%
\begin{acknowledgements}
P.S. acknowledges support from the the project VEGA 2/0004/16 of
the Science Agency. S.G. and P.H. acknowledge the support from
grants 19-16890S and 19-17102S of the Czech Science Foundation
(GA\v{C}R). P.S., S.G., and P.H. acknowledge support from the Joint
Mobility Project SAV-AV\,\v{C}R-18-03 of Academy of Sciences of
the Czech Republic and Slovak Academy of Sciences. 
J.M.J. thanks the STFC for support via funding 
given in his PhD studentship. D.M.L acknowledges 
support from the European Commission's H2020 Programme under the 
following Grant Agreements: GREST (no.~653982) and Pre-EST (no.~739500) 
and is grateful to the Science Technology and Facilities Council 
for the award of an Ernest Rutherford Fellowship (ST/R003246/1). 
This work utilises data obtained by the Global Oscillation Network 
Group (GONG) programme, managed by the National Solar Observatory, 
which is operated by AURA, Inc. under a cooperative agreement with 
the National Science Foundation. We are also thankful to anonymous
referee for his/her useful comments and suggestions to the paper. 
\end{acknowledgements}
%
%

%
\end{document}